\documentclass[preprint,10pt]{aastex}
\pdfoutput=1
\usepackage{grffile}  

\usepackage{url}                                                                                  

\renewcommand{\S}{Section }
\newcommand{\anchorfoot}[2] {\anchor{#1}{#2}\footnote{\url{#1}}}

\newcommand{\todo}[1]       {[[{\bf \footnotesize #1}]]} 

\newcommand{\tbr}[1]        {#1}

\newcommand{\revise}[1]     {{#1}}

\newcommand{\Spitzer} {{\em Spitzer}}
\newcommand{\Chandra} {{\em Chandra}}
\newcommand{\ACIS}    {{ACIS}}

\newcounter{column_number}
\newcommand{\numberthecolumn}{\colhead{(\arabic{column_number})}\stepcounter{column_number}}

\shorttitle{X-ray Source Classifier}
\shortauthors{Broos et al.}                                

\begin{document}



\title{A Naive Bayes Source Classifier for X-ray Sources}

\author{
Patrick S. Broos\altaffilmark{1},
Konstantin V. Getman\altaffilmark{1}, 
Matthew S. Povich\altaffilmark{1}\altaffilmark{2},
Leisa K. Townsley\altaffilmark{1}, 
Eric D. Feigelson\altaffilmark{1}, 
Gordon P. Garmire\altaffilmark{1} 
}
\email{patb@astro.psu.edu}

\altaffiltext{1}{Department of Astronomy \& Astrophysics, 525 Davey Laboratory, 
Pennsylvania State University, University Park, PA 16802, USA} 

\altaffiltext{2}{NSF Astronomy and Astrophysics Postdoctoral Fellow}

\begin{abstract}
The Chandra Carina Complex Project (CCCP) provides a sensitive X-ray survey of a nearby starburst region over $>$1 square degree in extent.  Thousands of faint X-ray sources are found, many concentrated into rich young stellar clusters.  However, significant contamination from unrelated Galactic and extragalactic sources is present in the X-ray catalog.  
We describe the use of a naive Bayes classifier to assign membership probabilities to individual sources, based on source location, X-ray properties, and visual/infrared properties.
For the particular membership decision rule adopted, 75\% of CCCP sources are classified as members, 11\% are classified as contaminants, and 14\% remain unclassified.
The resulting sample of stars likely to be Carina members is used in several other studies, which appear in a \anchor{TBD}{Special Issue} of the \apjs\ devoted to the CCCP.
\end{abstract}


\section{INTRODUCTION \label{intro.sec}}

A common stage in studies of astronomical populations is the division of samples into distinct classes. 
Often this is accomplished using measurements of a few similar quantities, such as visual or infrared photometric color indices.  
Simple decision rules are typically used, such as the localization of protostars defined by a box in $[5.8]-[8.0]$ vs.\ $[3.6]-[4.5]$ color-color diagrams \citep{Allen04}.  
But classification becomes more complex when the measured quantities have incompatible units or character, such as cases where both real-valued quantities and discrete quantities are available.  
A large field of applied mathematics---falling under the rubrics of supervised multivariate classification, data mining, pattern classification, and machine learning---has emerged to treat such problems.  

We confront this problem at an early stage of understanding the \Chandra\ Carina Complex Project (CCCP), which is devoted to the study of the nearest starburst region of our Galactic plane \citep{Townsley11} using X-ray observations with the Advanced CCD Imaging Spectrometer (\ACIS) on the {\em Chandra X-ray Observatory}.
More than 14,000 X-ray sources are identified by the CCCP, most of which clearly lie in young stellar clusters \citep{Broos11}.
A study of potential CCCP contaminants \citep{Getman11} predicts ${\sim}$5,200 of these sources will be unassociated with the Carina Complex: ${\sim}$2,500 extragalactic sources, ${\sim}$1,800 foreground Galactic field stars, and ${\sim}$900 background Galactic field stars.  

The goal of classification in the CCCP is to assign {\em individual} sources to one of four classes---foreground stars, Carina pre-main sequence stars, background stars, and extragalactic sources---so that subsequent studies can concentrate on the Carina members.
Equally important is to identify sources for which the available evidence does not strongly favor any of the classes; these sources will be designated as ``unclassified''.
The principle measured quantities powering the classification are the individual source X-ray properties, visual and infrared counterpart properties, and projected position with respect to known clusters.

We choose to implement a ``Naive Bayes Classifier'', one of the simpler classification methods developed in the fields of machine learning and pattern recognition \citep{Ripley96, Hastie01, Duda02}. 
An particularly accessible presentation of this intuitive method is provided by \anchorfoot{http://en.wikipedia.org/wiki/Naive_Bayes_classifier}{Wikipedia.}
For each observational quantity expected to carry classification information, a prediction for the distribution of that quantity expected from members of each class is constructed.
A prediction for the expected fractionation of the observed source catalog into the possible classes is also made, prior to examination of the measurements.
Class probabilities are then assigned to individual sources in a way that rationally combines our prior expectations regarding the odds for encountering each class with the consistency we find between the observations of that source and the predictions for each class.
Such a combination of ``prior'' belief and ``likelihood'' of the observed data is one of the fundamental attractions of all Bayes procedures.

\S\ref{clues.sec} describes the CCCP observational quantities of interest in more detail and the classification methodology is presented in \S\ref{strategy.sec}.  
The results of the classifier are shown in \S\ref{implementation.sec} with a validation effort outlined in \S\ref{validation.sec}.  
Quantitative classification results for the CCCP sources are given in an electronic table associated with this paper.

\section{CHARACTERISTICS OF CARINA MEMBERS AND CONTAMINATING POPULATIONS \label{clues.sec}}

We briefly list the properties, and associated measured quantities, useful for discrimination between Carina members and contaminant populations.  For each CCCP source, often only a subset of the observable properties listed below is available to influence our opinion as to that source's classification.

\revise{Below, we refer several times to the CCCP study by \citet{Getman11} of various populations of X-ray sources unrelated to Carina that are projected onto the field and are expected to contaminate the CCCP catalog.
Those authors performed detailed Monte Carlo simulations of extragalactic objects (primarily AGNs) seen through the Galactic disk and Galactic field stars (main sequence and giants from types F to M) distributed throughout the disk.
Their goals were to estimate how many objects in each population may be detected by the CCCP, and more importantly, to estimate the expected distribution of key observable quantities for those detected populations.
These simulations employed population synthesis models, published luminosity functions and spectral distributions, spatially-varying absorption across the CCCP field, the response of \Chandra, and the position-dependent sensitivity of the CCCP observations.
}

\begin{figure}[htb]
\centering
\includegraphics[width=0.5\textwidth]{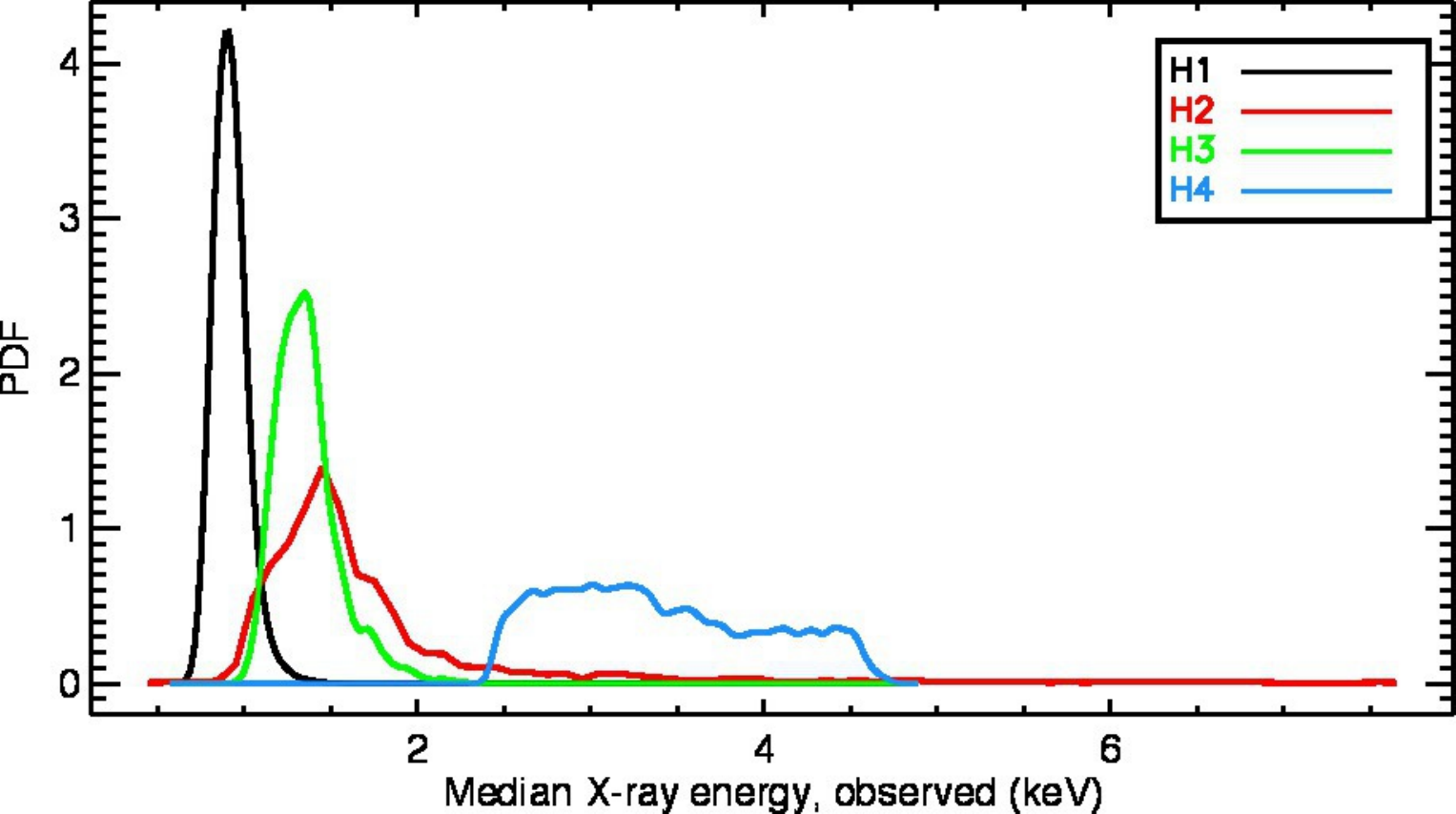}\\
\includegraphics[width=0.5\textwidth]{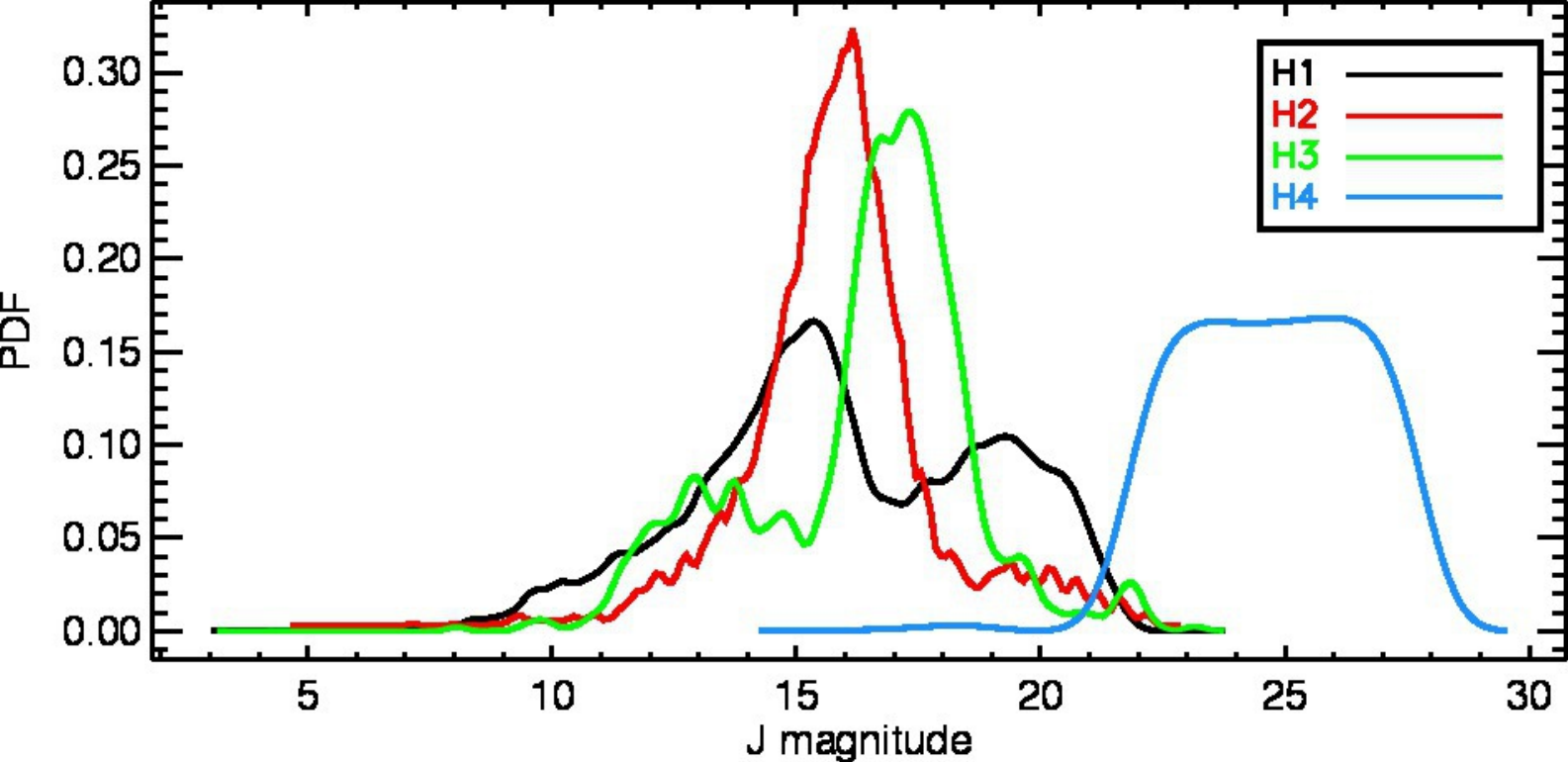}
\caption{Model distributions for $J$-band flux (upper panel) and apparent median X-ray energy (lower panel) for foreground stars (H1), young star in the Carina complex (H2), Galactic background stars (H3), and AGN (H4).
\label{evidence_distributions.fig}}
\end{figure}

\begin{description}

\item [Spatial location] \citet[][Fig.~1]{Getman11} find that the contaminating populations are nearly uniformly distributed across the CCCP field.
In contrast, the CCCP field has many historically well-known clusters \citep{Feigelson11} that must produce local over-densities of X-ray sources.
Clearly, before considering physical properties, the odds that a specific source is a Carina member increase with the local density of sources above the background level.

\item [Spectral types] About 200 visually-bright sources in the CCCP field have spectroscopically-determined spectral types \citep{Gagne11}.
Stars with OB spectral types, which are expected to be rare in the Galactic field population, can be reasonably assumed to lie in the Carina complex, whereas low-mass spectral types may be either Carina members or foreground main sequence stars.

\item [Magnetic activity] High amplitude, rapid X-ray flares are frequent in young stars but are less common in older stars \citep{Wolk05}.  
While nearly all extragalactic sources exhibit variability on timescales of days to months in the \Chandra\ band, only a small fraction ($<15$\%) have detectable variations within 1 day \citep{Paolillo04, Shemmer05}.  

\item [Apparent X-ray spectrum] Foreground stars should generally have lower line-of-sight absorption than Carina members, whereas background stars and extragalactic sources should have higher absorption.
The nature of the emitting plasma (e.g., temperature) is also expected to differ among these populations.
In combination, these two factors produce distinct shapes for the typical apparent X-ray spectra from these populations, which can be characterized (even for weak sources) by distinct distributions for the median X-ray energy statistic \citep[][Fig.~2]{Getman11} as shown in the upper panel of Figure~\ref{evidence_distributions.fig}.

\item [Apparent near-infrared brightness]  The distributions of $J$-band flux from simulations of the three contaminating populations by \citet[][Fig.~3]{Getman11} are shown in the lower panel of Figure~\ref{evidence_distributions.fig}.
Extragalactic sources are readily discriminated by their extremely faint $J$-band flux, due partly to absorption by dust in the Galactic mid-plane.    
Foreground field stars show a bimodal distribution from solar-type stars at moderate distances and nearby M dwarfs. 

\item [Apparent mid-infrared brightness] AGN (and point-like extragalactic sources in general) are observed to be faint in the mid-IR relative to young stars in nearby Galactic clusters \citep{Harvey07}. 
In the case of Carina, where extragalactic sources are observed through the absorption produced by the Galactic mid-plane, we expect AGN to appear fainter still.

\item [Circumstellar dust] Circumstellar dust in disks or infalling envelopes associated with young pre-main sequence stars produce near-infrared (NIR) and/or mid-infrared (MIR) emission in excess of a normally reddened stellar photosphere \citep[e.g.,][]{Meyer97,Povich11}.

\end{description}

\section{CLASSIFICATION STRATEGIES \label{strategy.sec}}

\subsection{Framework for Inferring Source Classification}

In order to infer a source's classification (or to conclude that it is undetermined), a clearly-defined framework for combining whatever observable data are available is required.
Note that the observable data may be in conflict for a particular source; for example, the source may have a median X-ray energy most consistent with a young star in Carina but have a $J$-band flux most consistent with an AGN.

One intuitive framework is a simple decision tree, an ordered network of simple decisions involving the observed data that leads us to a classification decision, or a conclusion that the classification is undetermined.
In such a system, real-valued observables (e.g., clustering, median X-ray energy, $J$-band flux) have to be compared with thresholds in order to construct nodes in the decision network.
Choosing such thresholds is a daunting and contentious task, although probabilistic decision tree construction techniques could be tried \citep[][Chpt. 8]{Duda02}.  

A more flexible framework introduces metrics---functions of the observations---into discrimination decisions.
Such metrics typically involve the offset of an observed datum from reference values that are typical for each class.  
Metrics can be weighted so that certain source properties trump others; for example, one might believe that detection of X-ray variability should be a more definitive classifier than $J$-band photometry.

A third approach, which we adopt here, is based on the intuition of maximizing the likelihood of the observed source data with respect to membership in different classes.
In this context, a likelihood is simply a joint probability distribution of all the observed data, conditioned on the class to which the source truly belongs,
  \begin{eqnarray}
  p(D_1,D_2, \ldots D_N \,|\, {\rm class}=H) \label{joint_likelihood.eqn}
  \end{eqnarray}
where $D_1,D_2, \ldots D_N$ are the $N$ observed quantities potentially available for each source.  $H$ is the set of mutually exclusive hypotheses,
  \begin{description}
  \item[H1:] source is a foreground Galactic field star
  \item[H2:] source is a young star in the Carina complex
  \item[H3:] source is a background Galactic field star
  \item[H4:] source is an extragalactic source.
  \end{description}

The likelihood function $p(D_1,D_2, \ldots D_N \,|\, {\rm class}=H)$ is a set of four {\em models}
  \begin{eqnarray}
  p(D_1,D_2, \ldots D_N \,|\, {\rm class}=H1) \nonumber \\
  p(D_1,D_2, \ldots D_N \,|\, {\rm class}=H2) \nonumber \\
  p(D_1,D_2, \ldots D_N \,|\, {\rm class}=H3) \nonumber \\
  p(D_1,D_2, \ldots D_N \,|\, {\rm class}=H4)
  \end{eqnarray}
that encode our understanding of what combinations of observable data are produced by sources from each of our four parent populations.
In principle, these models encapsulate our understanding of the physics of the X-ray and infrared emission of these four classes of objects, plus observational effects such as absorption, instrument response, and survey sensitivities.
Formally, these models are $N$-dimensional joint probability distributions that represent the many physical correlations that exist among the observable source properties.
Significant assumptions and informed professional judgments must be introduced to avoid an intractable problem.

We do not view a likelihood approach to the classification problem as inherently superior to a decision tree or set of weighted metrics.
However, we find that the likelihood approach helps us to clarify and quantify our beliefs regarding the classification conclusions that should be drawn from the various clues provided by the observations (\S\ref{clues.sec}), and provides a quantitative means for combining that evidence to make classification decisions.

\subsection{Naive Bayes Classifier} \label{naiveBayes.sec}

\subsubsection{Likelihoods \label{likelihoods.sec}}

We make the common and critical simplification that the joint likelihood in Equation~\ref{joint_likelihood.eqn} can be approximated by the product of one-dimensional likelihoods.
More formally, we assume that the observed properties of a source are statistically independent:
  \begin{eqnarray}
  p(D_1,D_2, \ldots D_N \,|\, {\rm class}=H) &=& p(D_1 \,|\, {\rm class}=H) \; p(D_2 \,|\, {\rm class}=H) \ldots p(D_N \,|\, {\rm class}=H) \nonumber \\
                                       &=& \prod_{i=1}^N p(D_i \,|\, {\rm class}=H).   \label{independent_likelihoods.eqn}
  \end{eqnarray}
The terms on the right side of Equation~\ref{independent_likelihoods.eqn} have a straightforward interpretation.
Each term represents four one-dimensional probability distributions, each representing the distribution of a particular observable property that we expect from each of the four possible source classes.
These likelihoods constitute astrophysical models that predict observable quantities for each of the four source populations, and are discussed in more detail in \S\ref{implementation.sec}.

For a particular CCCP source, the data $D_1,D_2, \ldots D_N$ have specific values obtained from observations.
The likelihood should be viewed as a function of H, the source class, which is a discrete variable with four possible values.
Thus, the observed data for each source produce four likelihood values defined by Equation~\ref{independent_likelihoods.eqn}.
The class with the largest likelihood is the maximum likelihood classification of the source.  

The assumption of independence is not strictly correct in the X-ray source classification problem.  
For example, a higher value of median X-ray energy will be somewhat correlated with fainter $J$-band flux, as both are products of heavier obscuration.  
Also, for Carina members, the detection of rapid X-ray variability will be correlated with X-ray flux, which itself is linked to pre-main sequence stellar mass and $J$ magnitude \citep{Telleschi07}.  

As with many multivariate problems, we must treat the case of missing data.
When an estimate of a property is not available for a specific X-ray source, we simply choose to omit that term from Equation~\ref{independent_likelihoods.eqn}.
The missing source property thus plays no role in the classification decision\footnote{
In principle, an alternative policy for missing data could be adopted.
With enough modeling of the data collection process, one could choose to interpret the absence of a datum as an upper or lower limit on its value, and then use that limit in the likelihood calculation.
We have chosen not to add this complexity.}.

\subsubsection{Prior Class Probabilities \label{priors.sec}}

Suppose a source were selected from the CCCP catalog, we were told nothing about it whatsoever, and we were asked to assign odds to each of the four classes from which it could have been drawn.
The obvious odds we would assign to each contaminant class would be the ratio of the total number of that contaminant predicted by \citet{Getman11} to the size of the catalog; the odds assigned to the H2 class (member) would then be calculated such that the four odds sum to unity.
Classification probabilities such as these---estimated prior to examining the measured properties of individual sources---\revise{are commonly referred to} as {\em prior probabilities}, or ``priors'' informally.

Our classification problem can be formulated in terms of a single set of prior classification probabilities, applicable to all sources in the catalog, as described above.  
Under this formulation, the observed position of each source is a measurement (akin to median X-ray energy, $J$-band flux, etc.), and thus has a likelihood term in Equation~\ref{independent_likelihoods.eqn}, as discussed in the Appendix.
However, we find it convenient to adopt a different and equivalent formulation in which separate prior classification probabilities are computed for each source, based on the local density of detected sources and \revise{on} the local densities of expected contaminants.
More formally, for each location $\mathbf{r}$ on the sky at which a source is detected we compute a discrete prior distribution for the source class $H \in \{H1,H2,H3,H4\}$, using the expected \revise{local} densities of the three contaminants, $\rho_{H1}(\mathbf{r}),\rho_{H3}(\mathbf{r}),\rho_{H4}(\mathbf{r})$, and an estimate of the observed source density, $\rho_{\rm obs}(\mathbf{r})$:
  \begin{eqnarray}
    {\rm prior}(\mathbf{r})_{H1} =&  \rho_{H1} &/ \rho_{\rm obs}(\mathbf{r}) \nonumber \\
    {\rm prior}(\mathbf{r})_{H2} =& (\rho_{\rm obs}(\mathbf{r}) - \rho_{H1} - \rho_{H3} - \rho_{H4}) 
                            \label{prior2.eqn} &/ \rho_{\rm obs}(\mathbf{r}) = 1 - \frac{\rho_{H1}+\rho_{H3}+\rho_{H4}}{\rho_{\rm obs}(\mathbf{r})} \\ 
    {\rm prior}(\mathbf{r})_{H3} =&  \rho_{H3} &/ \rho_{\rm obs}(\mathbf{r}) \nonumber\\
    {\rm prior}(\mathbf{r})_{H4} =&  \rho_{H4} &/ \rho_{\rm obs}(\mathbf{r}).\nonumber
  \end{eqnarray}  
In this study, the spatial distributions of the contaminants are expected to be nearly uniform, so $\rho_{H1},\rho_{H3},$ and $\rho_{H4}$ appear above as scalars rather than as position-dependent quantities.
\revise{The normalizing constant $\rho_{\rm obs}(\mathbf{r})$ is required so that $\sum_{H\in\{H1,H2,H3,H4\}} {\rm prior}(\mathbf{r}) = 1$.}

\begin{figure}[htb]
\centering
\includegraphics[width=0.5\textwidth]{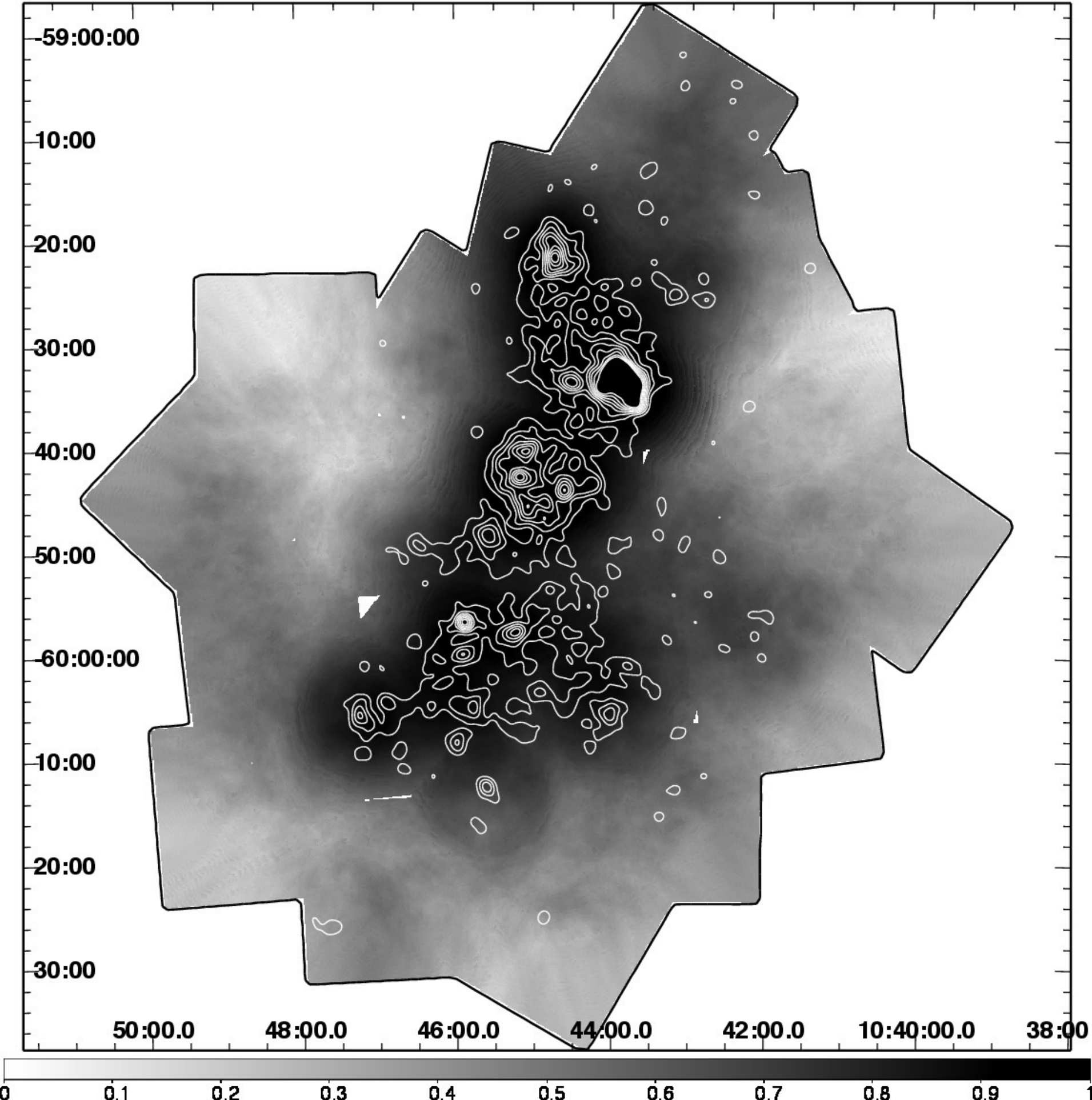}
\caption{\revise{Spatial prior probability for the H2 class (grayscale map) estimated from the observed source density, $\rho_{\rm obs}(\mathbf{r})$, (shown as contours) via Equation~\ref{prior2.eqn}:} 
$ {\rm prior}(\mathbf{r})_{H2} = 1 - \frac{\rho_{H1}+\rho_{H3}+\rho_{H4}}{\rho_{\rm obs}(\mathbf{r})}. $ 
\label{H2_prior.fig}}
\end{figure}

The spatial prior map for the H2 class, ${\rm prior}(\mathbf{r})_{H2}$, representing the over-density of observed sources above the expected density of contaminants, is shown in Figure~\ref{H2_prior.fig}. 
In this study the minimum value of the H2 prior across the field was positive (0.06); if Equation~\ref{prior2.eqn} had produced negative or very small values (i.e., the observed source density had been less than the predicted contaminant density) at any location, then we would have imposed a floor on the H2 prior and adjusted the equations above so that the four prior probabilities summed to unity at every location.

\subsubsection{Posterior Probabilities}

Bayes' Theorem provides a conceptually simple and principled method for combining likelihoods computed from data (\S\ref{likelihoods.sec}) and prior probabilities for model parameters (\S\ref{priors.sec}) to produce {\em posterior probabilities} for the model parameters, conditioned on the data we have observed:
  \begin{eqnarray}
  p(H \,|\, D_1,D_2, \ldots D_N) & \propto & p(D_1,D_2, \ldots D_N \,|\, H) \; p(H) \nonumber \\
  \mbox{OR~~~~~~~posterior}                  & \propto &  \mbox{likelihood} \times \mbox{prior.} \label{BT.eqn}
  \end{eqnarray}
In our classification task we have one model parameter, the discrete classification hypothesis $H$ for a source at location $\mathbf{r}$ with observed source properties $D_1,D_2, \ldots D_N$.
The posterior probabilities for the four possible values of $H$ can be written as
  \begin{eqnarray}
  {\rm Prob}({\rm class}=H1 \,|\, \mathbf{r},D_1,D_2, \ldots D_N) \propto \prod_{i=1}^N p(D_i \,|\, {\rm class}=H1) \, {\rm prior}(\mathbf{r})_{H1} \nonumber  \\
  {\rm Prob}({\rm class}=H2 \,|\, \mathbf{r},D_1,D_2, \ldots D_N) \propto \prod_{i=1}^N p(D_i \,|\, {\rm class}=H2) \, {\rm prior}(\mathbf{r})_{H2} \nonumber \\
  {\rm Prob}({\rm class}=H3 \,|\, \mathbf{r},D_1,D_2, \ldots D_N) \propto \prod_{i=1}^N p(D_i \,|\, {\rm class}=H3) \, {\rm prior}(\mathbf{r})_{H3} \nonumber \\
  {\rm Prob}({\rm class}=H4 \,|\, \mathbf{r},D_1,D_2, \ldots D_N) \propto \prod_{i=1}^N p(D_i \,|\, {\rm class}=H4) \, {\rm prior}(\mathbf{r})_{H4} \label{posterior4.eqn} 
  \end{eqnarray}
The common constant of proportionality in these equations is easily found by requiring that the four posterior probabilities sum to unity, after which they can be interpreted as probabilities.

The classification system we have described is one of the simplest probabilistic procedures for multivariate classification, is widely used, and is commonly referred to as a Naive Bayes Classifier. The term ``naive'' refers to the assumption of statistical independence in deriving the likelihood.  The naive Bayes classifier is known to perform well in many cases with mild violations of independence \citep{Hand01}.  The method is described within the context of other procedures for supervised classification of multivariate datasets in textbooks by \citet{Ripley96}, \citet{Hastie01}, and \citet{Duda02}.     
Recent applications to astronomical problems include works by \citet{Bazell01}, \citet[][Appendix A]{Norman04}, \citet{Zhang04}, \citet{Picaud05}, \citet[][Appendix A]{Ptak07}, \citet{Mahabal08a}, \citet{Mahabal08b}, and \citet{Burnett10}.

\section{IMPLEMENTATION FOR CARINA SOURCES \label{implementation.sec}}

\subsection{Assigning the Likelihoods}

\citet{Broos11} and citations therein describe the X-ray, visual, NIR, and MIR measurements available for CCCP sources.
From those data, eight observed or physical source properties were derived.
Four likelihood functions were assigned to each source property (Equation~\ref{independent_likelihoods.eqn}), representing the distribution of property values that we expect from members of each of the four possible source classes.
These eight properties and their likelihoods are described below.

For three properties (median X-ray energy, $J$-band flux, and X-ray variability) no estimate of the distribution expected from Carina members (the H2 likelihood function) was readily available {\it a priori}.
These distributions were estimated empirically, using a training set consisting of likely H2 sources within the field of view of the deep HAWK-I infrared survey, which contains all the large clusters in Carina \citep[][]{Preibisch11}.
This region on the sky has the most counterpart information available and has a small fraction of contaminants.
The classifier itself was used iteratively to select the training set within this region.
Initially, H2 likelihoods for these three properties were not available, and the classifier evaluated membership using only the other five likelihoods and the spatial priors.
Once an initial training set of likely H2 sources was identified, classification was re-evaluated (using all eight properties) and the training set was re-defined, proceeding iteratively until a stable training set was defined.

\begin{description}
\item[Median X-ray energy:]  
The median X-ray energy \citep[MedianEnergy\_t in][]{Broos11} distributions assumed for the four classes are shown in Figure~\ref{evidence_distributions.fig} (upper panel).
The contaminant distributions (H1,H3,H4) were obtained from simulations by \citet{Getman11}.
The Carina member distribution (H2) was obtained iteratively from the training set.

\item[$J$-band flux:]  
$J$-band photometry was obtained from the SOFI, HAWK-I, and 2MASS catalogs \citep{Broos11}.
The distributions assumed for the four classes are shown in Figure~\ref{evidence_distributions.fig} (lower panel).
The contaminant distributions (H1,H3,H4) were obtained from simulations by \citet{Getman11}.
The Carina member distribution (H2) was obtained iteratively from the training set.

\item[X-ray variability:]  
In the X-ray catalog, variability is quantified by a p-value\footnote
{
In statistical hypothesis testing, the p-value is the probability of obtaining a test statistic (such as the Kolmogorov-Smirnov statistic) at least as extreme as the one that was actually observed when the null hypothesis is true.
}
for the no-variability hypothesis, estimated via the Kolmogorov-Smirnov (KS) statistic \citep[ProbKS\_single in][]{Broos11}.
In the classifier, this p-value is discretized into a variability grade with three values, defined in Table~\ref{variability.tbl}.
Since the contaminant populations (H1,H3,H4) are assumed to be constant on relevant timescales (the null hypothesis), the expected distribution of this grade is by definition obtained from the p-values used to define the flag, as shown in Table~\ref{variability.tbl}. 
For example, the ``definitely variable'' grade is assigned when a p-value $< 0.005$ is found for the KS statistic; for constant sources  this will occur by chance with a probability of 0.005, and thus the likelihoods for the ``definitely variable'' observation are assigned that value for the H1, H3, and H4 classes.
The Carina member distribution (H2) was obtained iteratively from the training set.

Note that the probabilities shown in Table~\ref{variability.tbl} define proper likelihood functions; for each class hypothesis the sum of the probabilities for all possible observational outcomes (rows) is unity.
Because the posterior class probabilities in Equation~\ref{posterior4.eqn} are normalized to sum to unity, the influence a particular observation has on classification depends on the relative, not the absolute, likelihoods for that observation across the classes.
For example, a variability grade of ``definitely variable'' is interpreted as strong evidence for the H2 class because the H2 likelihood is ${\sim}$15 times larger (0.076/0.005) than the other likelihoods.
The grade ``possibly variable'', with a likelihood ratio of ${\sim}$4 (0.10/0.045), is interpreted as moderate evidence for the H2 class.
Inevitably, the finding of ``no evidence'' for variability is interpreted as weak evidence {\em against} the H2 class because the likelihood ratio (0.82/0.95) is less than unity.
This logical, but unintuitive inference might be overlooked in an informal classification framework based on decision trees or metrics.


\begin{deluxetable}{rcr@{}l@{}lll}
\tablecaption{Likelihood functions for variability grades  \label{variability.tbl}
}
\tablewidth{0pt}
\tabletypesize{\footnotesize}
\tablehead{
\colhead{Grade} & &  \multicolumn{3}{c}{Definition}               & \multicolumn{2}{c}{${\rm Prob}(D_3 \,|\, H)$} \\ \cline{6-7}%
                & &    &&                                         & \colhead{H1,H3,H4} & \colhead{H2}
}
\startdata
No evidence for variability & $\Leftrightarrow$ & 0.05\phn $<$~ & ProbKS\_single &             & 1.00\phn-\phn0.05        & 0.82  \\
          Possibly variable & $\Leftrightarrow$ & 0.005    $<$~ & ProbKS\_single & ~$<$ 0.05   & 0.05\phn-\phn0.005       & 0.10  \\
        Definitely variable & $\Leftrightarrow$ &               & ProbKS\_single & ~$<$ 0.005  & 0.005 & 0.076
\enddata
\end{deluxetable}                                            

\item[Visual spectroscopy:]  
Spectral type estimates obtained from visual spectroscopy are available only for a few sources with high apparent brightness \citep{Broos11}, but are believed to be reliable.
We interpreted spectral types AFGKM as conclusive evidence that the source is in the foreground, and spectral types OB as conclusive evidence the source is a Carina member.
That policy is represented by the likelihood functions in the $D_4$ section of Table~\ref{likelihoods.tbl}.

\begin{deluxetable}{rrrrr}
\tablecaption{Likelihood functions for source properties $D_4 \ldots D_8$ \label{likelihoods.tbl}
}
\tablewidth{0pt}
\tabletypesize{\footnotesize}
\tablehead{
\colhead{Observed Property} & \multicolumn{4}{c}{${\rm Prob}(D_i \,|\, H)$} \\ \cline{2-5}%
                            & \colhead{H1} & \colhead{H2} & \colhead{H3} & \colhead{H4}
}
\startdata
\multicolumn{5}{l}{$D_4$: Spectral Type} \\
FGKM &  1 & 0 & 0 & 0  \\
OBA  &  0 & 1 & 0 & 0  \\
\hline
\multicolumn{5}{l}{$D_5$: [4.5 \micron]} \\
$<$ 13 mag &  1          & 1          & 1          & 0  \\
$>$ 13 mag &  \nodata    & \nodata    & \nodata    & \nodata \\
\hline
\multicolumn{5}{l}{$D_6$: MIR Colors} \\
Infrared excess         &  0 & 1 & 0 & 0  \\
No infrared excess      &  \nodata & \nodata & \nodata & \nodata \\ 
\hline
\multicolumn{5}{l}{$D_7$: SED Model} \\
Infrared excess    &  0.03 & 0.90 & 0.03 & 0.03  \\
No infrared excess &  \nodata & \nodata & \nodata & \nodata \\ 
\hline
\multicolumn{5}{l}{$D_8$: NIR Colors} \\
Infrared excess    &  0.03 & 0.90 & 0.03 & 0.03  \\
No infrared excess &  \nodata & \nodata & \nodata & \nodata 
\enddata
\end{deluxetable}

%

\item[4.5 \micron\ photometry:] 
\Spitzer/IRAC photometry was obtained from the SpSmith and SpVela catalogs \citep{Broos11}.
We do not use the 4.5 \micron\ flux value itself as a classification property (as was done for $J$-band flux) because we do not have models for the detailed 4.5 \micron\ flux distribution expected from our four classes.
Instead, we choose $[4.5] < 13$ mag as a definitive selection against H4 because the majority of extragalactic point sources detected by \Spitzer\ are observed to be fainter than this limiting magnitude \citep{Harvey07}. 
This criterion is conservative in that AGN observed through the obscuration produced by the Carina cloud and intervening segment of the Galactic plane are expected to be even fainter. 
We construct a bright/weak Boolean datum by applying a threshold of 13 magnitudes to the 4.5 \micron\ flux data using the likelihood functions shown in the $D_5$ section of Table~\ref{likelihoods.tbl}.
When weak 4.5 \micron\ flux is found, we choose to omit this property (second table row) because we cannot quantify how strongly that finding should be interpreted as weak evidence for the H4 class.

%

\item[Mid-infrared color-magnitude region:] 
\Spitzer/IRAC photometry was obtained from the SpSmith and SpVela catalogs \citep{Broos11}.
We adopt the color-magnitude selection described by \citet[][Equations (3) and (5)]{Robitaille08} for the \Spitzer\ GLIMPSE survey to identify intrinsically red mid-infrared stellar sources with dusty circumstellar material. 
Since this selection is very conservative and no contaminant population is expected to satisfy this property, we interpreted it as conclusive evidence of Carina membership (H2) as shown by the first row of the $D_6$ section of Table~\ref{likelihoods.tbl}.
As was the case with the bright 4.5 \micron\ flux property, we are unable to quantify how strongly the lack of a clear MIR excess should weigh against the H2 class, thus we chose to ignore this term in such cases (second row of the $D_6$ table).

%

\item[Infrared spectral energy distribution:]  
Spectral energy distribution (SED) modeling using near- and mid-infrared photometry can identify sources exhibiting a robust infrared excess best explained by circumstellar dust in a disk or infalling envelope \citep{Povich11}.
Photometry was obtained from the 2MASS and SpVela catalogs \citep{Broos11}.
We interpret such a detection of an infrared excess as strong but uncertain evidence of Carina membership (H2), as shown by the first row of the $D_7$ section of Table~\ref{likelihoods.tbl}.
\revise{We used our professional judgment to choose the strength of this evidence for H2, represented by the likelihood ratio 0.90/0.03.
}
As for some other source properties above, we are unable to quantify how strongly the lack of an infrared excess in SED fits should weigh against the H2 class so we chose to ignore this term in such cases (second row of the $D_7$ table).

%

\item[Near-infrared color-color region:]  
Although less frequent than in the mid-infrared, protoplanetary disk dust emission sometimes appears as a $K$-band excess in near-infrared color-color diagrams.
$JHK$-band color data were obtained from the SOFI and 2MASS catalogs \citep{Broos11}.
At the time this analysis was performed, the HAWK-I catalog exhibited small spatially varying systematic uncertainties in colors \citep{Preibisch11}; to simplify our analysis we omitted HAWK-I data here. 
For classification purposes, we define $K_s$-excess objects as those with $H-K$ colors that are $1.5 \sigma$ redder than the line defining reddened main sequence photospheres at mass 0.1~M$_\odot$.  
In addition, $J-H$ colors must be above the locus of classical T Tauri stars \citep{Meyer97}.  

We interpret detection of this $K_s$-band excess as strong but uncertain evidence of Carina membership (H2), as shown by the first row of the $D_8$ section of Table~\ref{likelihoods.tbl}.
\revise{As for the SED likelihood, the strength assigned to this evidence is based on our professional judgment.
}
Note that analysis of the infrared excess sources identified via SED fitting reveals that only a minority (${<}40\%$) would have been identified via $K_s$ excess \citep{Povich11}.
Lack of $K_s$ excess does not weigh against the H2 class, because stars with disks often do not separate cleanly from reddened, diskless stars in the JHKs color-color diagram; thus we chose to ignore this term in such cases (second row of the $D_8$ table).

%

\end{description}

\revise{Uncertainties on observed data tend to dull the effectiveness of any classifier.
One way to understand this effect is to imagine that the likelihoods (such as those in Figure~\ref{evidence_distributions.fig}) for the {\em measured} source properties (as opposed to the true values) will broaden and overlap more as the uncertainties rise.
An alternate way to represent data uncertainties is to define the likelihoods as distributions for {\em perfect} data, and then to {\em evaluate} the likelihood for a specific measurement by convolving the likelihood function with a distribution that represents the measurement (e.g., a Gaussian distribution with mean at the measured value and width equal to the estimated uncertainty on the measurement);  see for example \citet{Burnett10} and \citet{Norman04}.

In this study, our first foray into Bayesian methods, we have not attempted to formally represent data uncertainties.
The resulting distortion to the median X-ray energy likelihood was mitigated by ignoring the 3610 least reliable median energy measurements. 
Distortion to the J magnitude likelihood was limited by rejecting photometry values flagged as poor quality; the uncertainties on the ${\sim}8600$ J magnitudes used in the classification are very small (mean of 0.027 mag).
The KS statistic used to define an X-ray variability likelihood already accounts for the finite number of X-ray events observed.
The 4.5 \micron\ likelihood and the mid-infrared color-magnitude likelihood involve selections on MIR photometry that were defined using \Spitzer\ data, presumably with uncertainties comparable to our \Spitzer\ data.
The SED modeling process used to identify infrared excess and the NIR color space criteria used to identify $K$-band excess make use of cataloged photometry uncertainties.
}


\subsection{Posterior Class Probabilities}

\begin{figure}[htb]
\centering
\includegraphics[width=0.5\textwidth]{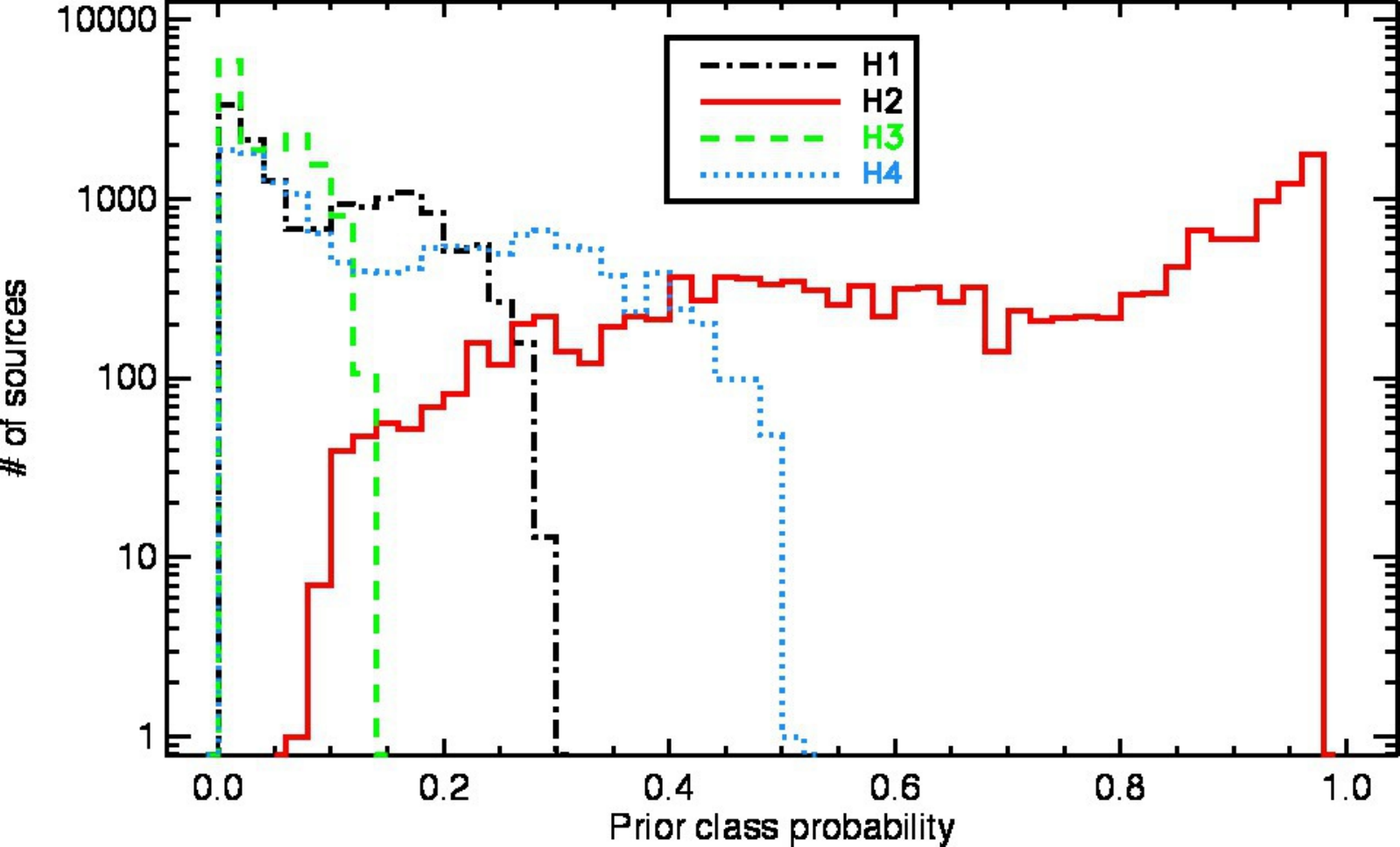}     \\
\includegraphics[width=0.5\textwidth]{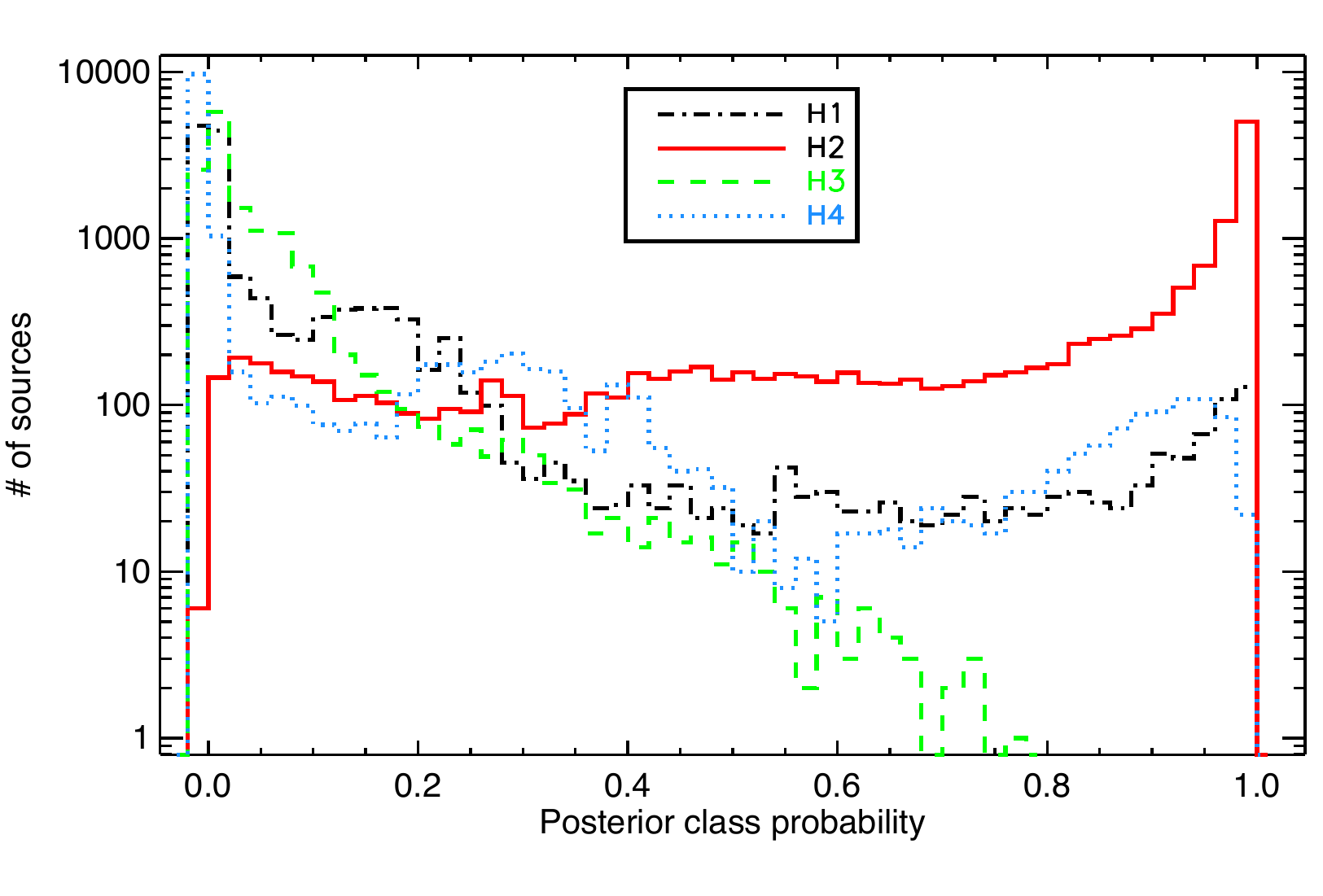}
\caption{Histograms of H1 (black, dash-dotted), H2 (red, solid), H3 (green, dashed), and H4 (blue, dotted) class probabilities for all CCCP sources, computed using only source position information (upper panel) and computed using both position information and data likelihoods (lower panel).
\label{prior_and_posterior.fig}}
\end{figure}

The source-specific priors described in \S\ref{priors.sec} (Equation~\ref{prior2.eqn}) represent the classification probabilities that we would adopt if no observations of source properties beyond position were available.
The upper panel of Figure~\ref{prior_and_posterior.fig} shows a histogram for each of these priors.
For example, a source appearing near the core of a dense cluster would appear near 1.0 in the histogram of the H2 prior (red), reflecting its near certain membership in the cluster, and would appear near zero in the other histograms (since the four classification probabilities sum to unity for each source).
Clearly, there are many such sources, and a classification based on source position alone would be useful.
In contrast, no sources are deemed likely to be contaminants based on only their position (i.e., the H1, H3, and H4 histograms are empty above ${\sim}$0.5).

The lower panel of Figure~\ref{prior_and_posterior.fig} shows posterior class probabilities for all CCCP sources inferred using both the spatial priors (upper panel) and the likelihoods for all observed data, via Equation~\ref{posterior4.eqn}.   
This graph is the principal result of the naive Bayes probability calculation given in Equation~\ref{posterior4.eqn} with likelihoods presented in Figure~\ref{evidence_distributions.fig} and Tables~\ref{variability.tbl} and \ref{likelihoods.tbl}.
Including the data likelihoods allows us to form more definite opinions about individual classifications, as can be seen by the larger number of sources in the lower panel with class probabilities near 1.0 or near zero compared to the upper panel.  
Now, some individual sources are identified as likely contaminants with large values in the H1, H3, and H4 histograms.
The H2 class is ruled out for a few sources with near-zero values in the H2 histogram.

\begin{figure}[htb]
\centering
\includegraphics[width=0.5\textwidth]{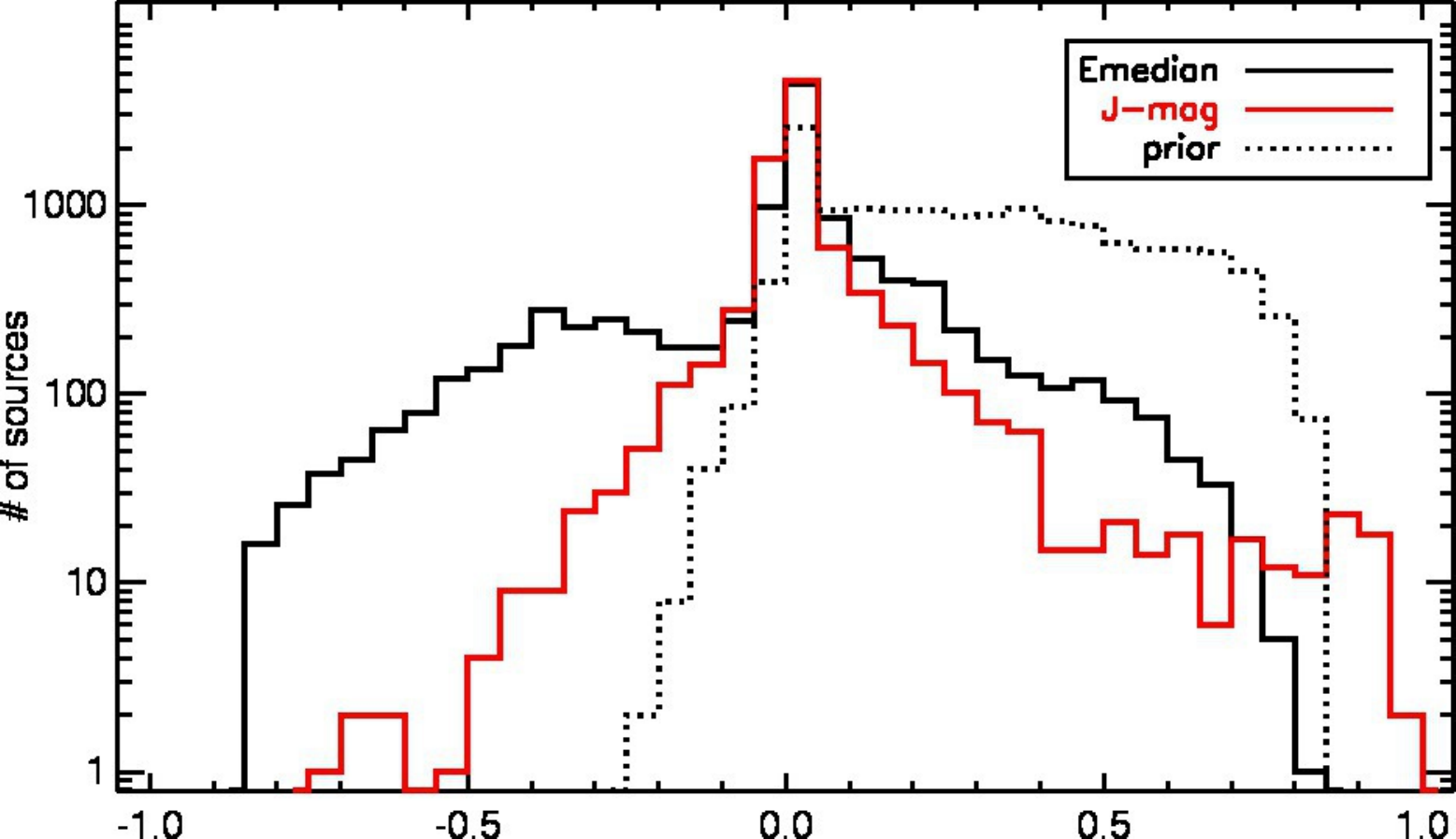}\\
\includegraphics[width=0.5\textwidth]{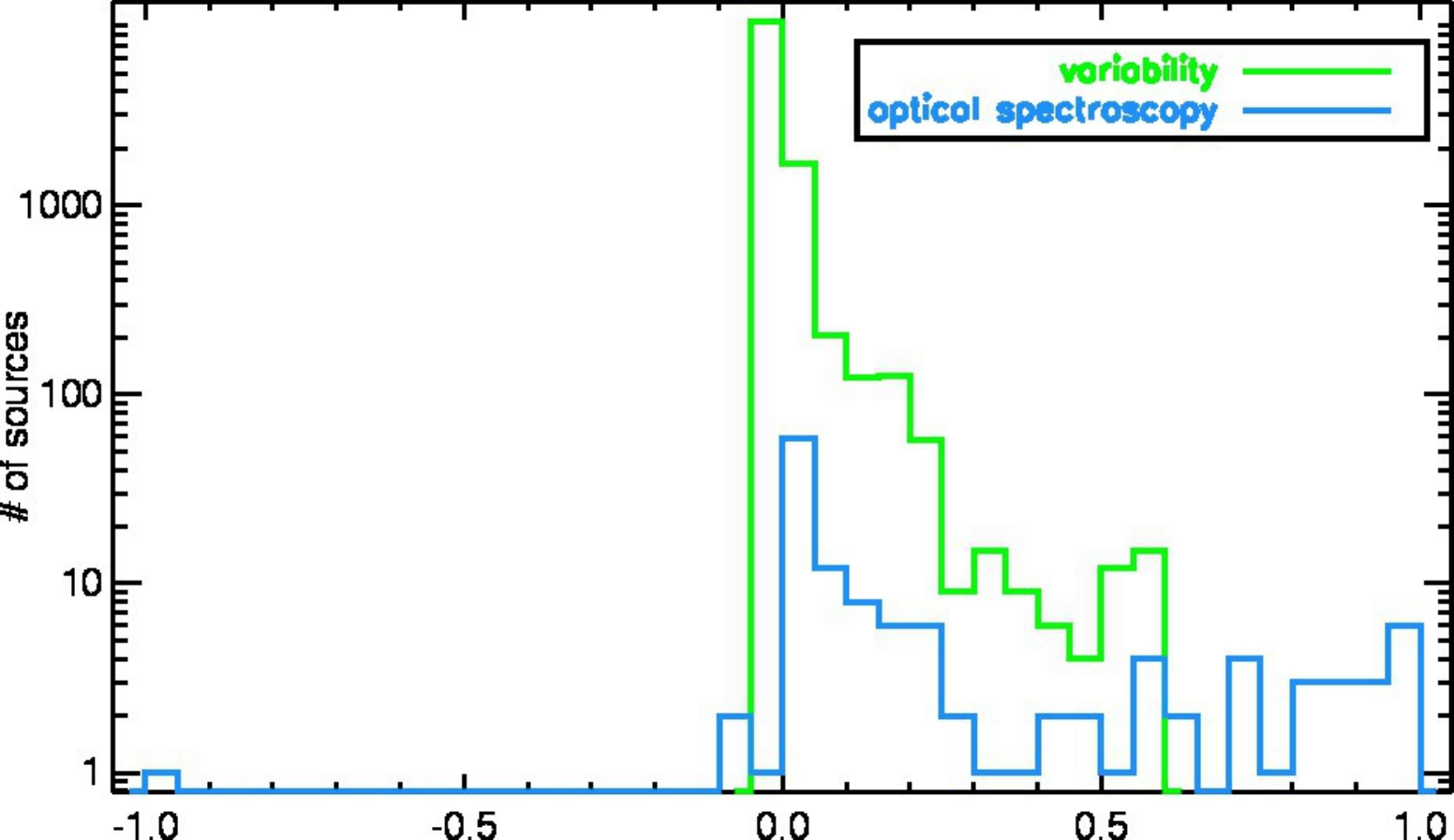}\\
\includegraphics[width=0.5\textwidth]{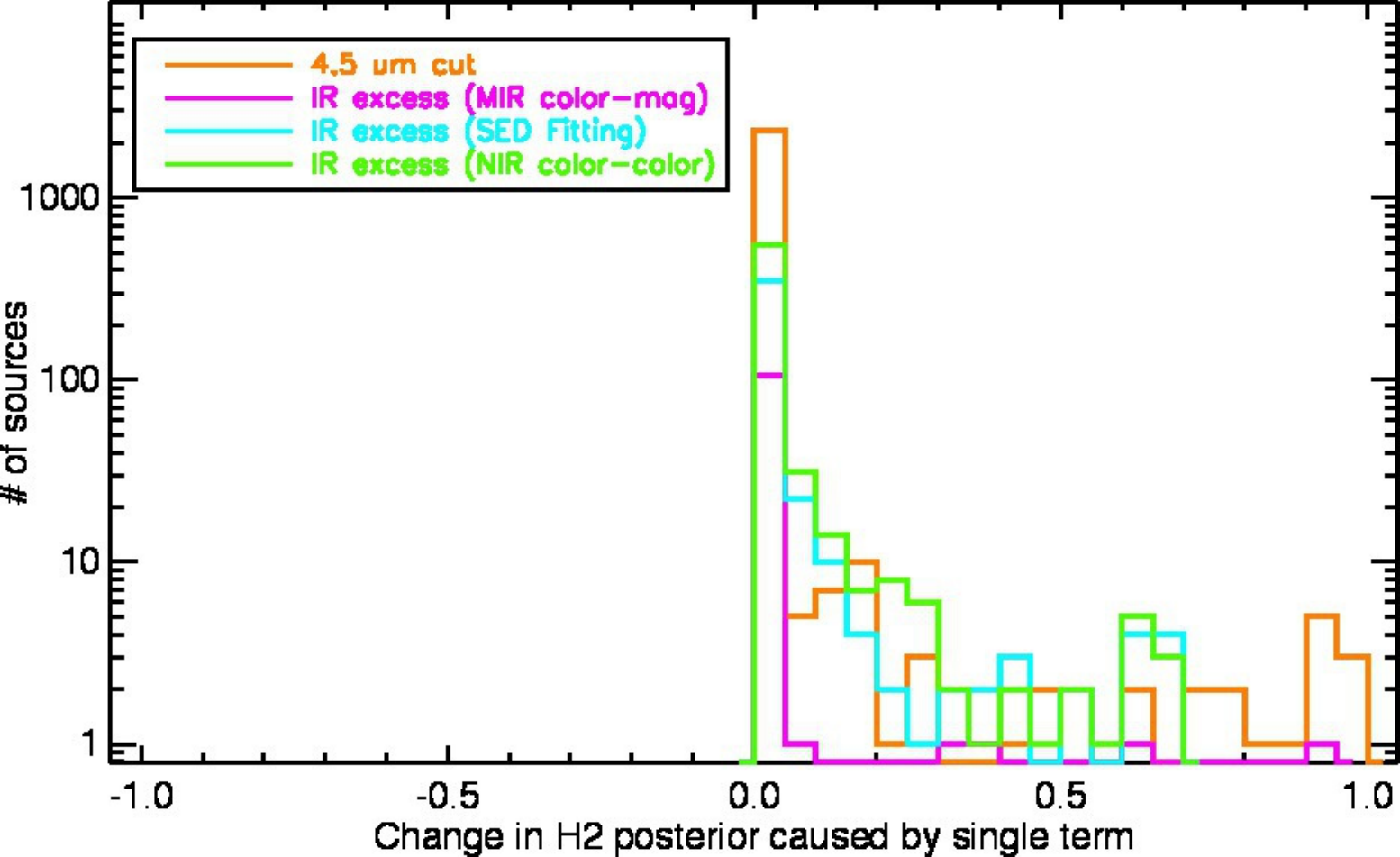}
\caption{Change in posterior probability for the H2 class (Carina member) when a single term (spatial prior or one likelihood) is omitted from the calculation.
\label{sensitivity.fig}}
\end{figure}

To better understand the influences that the spatial prior and source property likelihoods have on the posterior class probabilities, we repeated the classification calculation nine times, omitting a single term in Equation~\ref{posterior4.eqn} for each calculation.
Figure~\ref{sensitivity.fig} shows nine histograms, each depicting the change in the H2 posterior probability resulting from omission of the single specified term.
The upper panel shows three terms (the spatial prior, $J$-band flux, and median X-ray energy) that provide significant evidence both for (positive values) and against (negative values) the H2 classification.
The middle panel shows two terms (X-ray variability and visual spectroscopy) that provide significant evidence for (positive values) and minor evidence against (negative values) the H2 classification. 
The lower panel shows four terms (4.5 \micron\ photometry, MIR color-magnitude region, infrared SED models, and NIR color-color region) that, by construction, provide only evidence for the H2 classification. 
In the histograms corresponding to data likelihood terms, sources are omitted when data are not available.

Figure~\ref{sensitivity.fig} shows that spatial location (upper panel, dotted) strongly increases our belief in the H2 classification (Carina member) for thousands of CCCP sources. 
$J$-band flux and median X-ray energy (upper panel, black and red) are not decisive for most sources, but do provide key evidence both for and against the H2 class for a significant minority.
The other data likelihoods are available for fewer sources, and only rarely provide decisive evidence for Carina membership.

\revise{Bayesian posterior probabilities should be regarded as distinct from traditional hypothesis tests (such as goodness-of-fit tests and the KS test) in which the consistency between the observed data and a {\em single} hypothesis/model is quantified.
In hypothesis testing, finding a poor consistency between the data and model can rightly be interpreted as evidence that the model (null hypothesis) should be rejected; however finding good consistency does not in any sense prove the model (null hypothesis) is correct because  there may be many other models that well fit the data.

In contrast, the formulation of a Bayesian classifier articulates all reasonable {\em alternatives} to the hypothesis/model of interest (in our case H2=``cluster member''), and the {\em relative} consistency between the data and each of the models (class hypotheses) is evaluated (by the likelihood terms).
The absolute consistency between the data and a model---the concept measured by goodness-of-fit statistics and p-values---plays no explicit role in Bayesian inference.
For example, if the data are equally consistent with all the hypotheses/models---whether that consistency is very good or very bad---then the data will have no influence on our beliefs; the posterior class probabilities will be equal to the prior probabilities.
To the extent that all reasonable astrophysical classes of X-ray sources have been considered by our model, the resulting posterior probabilities should be interpreted as quantifying on an absolute scale the confidence with which we can infer that the source belongs to a particular class.
An additional advantage of Bayesian methods is that they provide a clear mechanism to incorporate {\it a priori} information about the quantities of interest.

Although posterior probabilities should be interpreted as our confidence in the four class
hypotheses, their {\em accuracy} will of course depend on the correctness of the likelihood
functions and class prior probabilities adopted, and on the choice of source properties considered.
In \S\ref{validation.sec} we assess the validity of our CCCP source classification.
}

\newpage
\subsection{An Example Calculation}

\revise{The calculation of posterior class probabilities for a hypothetical source is shown in Table~\ref{example.tbl}.
We imagine that this source lies at a random position in the Carina field, and thus we adopt an H2 class prior probability equal to the median (0.51) of the H2 prior map shown in Figure~\ref{H2_prior.fig}, and adopt the corresponding H1, H3, and H4 priors (0.15, 0.07, and 0.27).
These are shown in the top section of the table.
The eight source properties observed for this source, $D_1 \ldots D_8$, are shown in the middle table section; for simplicity only four of those produce likelihoods, ${\rm Prob}(D_i \,|\, H)$, in this example.
In each row of the table, boldface marks the class most favored by the prior, likelihood term, or posterior reported on that row. 
}

\begin{deluxetable}{llllll}
\tablecaption{Example Calculation of Posterior Probabilities \label{example.tbl}
}
\tablewidth{0pt}
\tabletypesize{\footnotesize}
\tablehead{
\colhead{} & \colhead{H1} & \colhead{H2} & \colhead{H3} & \colhead{H4} & {Note}
}
\startdata
{Class Priors}                                  & 0.15 & {\bf 0.51} & 0.07 & 0.27  & $\sum_{H=1}^4 = 1$ \\
\hline

{Observed Properties; Data Likelihoods} \\
\hspace*{1em}
$D_1$: Median X-ray Energy = 1.0 keV \\
\multicolumn{1}{r}{${\rm Prob}(D_1 \,|\, H) =$} & {\bf 2.52}  & 0.336 & 0.089 & 0.0 & Figure~\ref{evidence_distributions.fig} (upper panel) \\
\hspace*{1em}
$D_2$: [J] = 17.3 mag \\
\multicolumn{1}{r}{${\rm Prob}(D_2 \,|\, H) =$} & 0.071 & 0.103 & {\bf 0.279} & 0.002 & Figure~\ref{evidence_distributions.fig} (lower panel) \\
\hspace*{1em}
$D_3$: No evidence for variability \\
\multicolumn{1}{r}{${\rm Prob}(D_3 \,|\, H) =$} & {\bf 0.95}  & 0.82  & {\bf 0.95}  & {\bf 0.95} & Table~\ref{variability.tbl}\\
\hspace*{1em}
$D_4$: No spectral type \\
\multicolumn{1}{r}{${\rm Prob}(D_4 \,|\, H) =$} &\nodata&\nodata&\nodata&\nodata & Table~\ref{likelihoods.tbl} \\
\hspace*{1em}
$D_5$: [4.5 \micron] $>$ 13 mag \\
\multicolumn{1}{r}{${\rm Prob}(D_5 \,|\, H) =$} &\nodata&\nodata&\nodata&\nodata & Table~\ref{likelihoods.tbl} \\
\hspace*{1em}
$D_6$: No MIR excess \\
\multicolumn{1}{r}{${\rm Prob}(D_6 \,|\, H) =$} &\nodata&\nodata&\nodata&\nodata & Table~\ref{likelihoods.tbl} \\
\hspace*{1em}
$D_7$: No SED excess      \\
\multicolumn{1}{r}{${\rm Prob}(D_7 \,|\, H) =$} &\nodata&\nodata&\nodata&\nodata & Table~\ref{likelihoods.tbl} \\
\hspace*{1em}
$D_8$: No NIR excess \\
\multicolumn{1}{r}{${\rm Prob}(D_8 \,|\, H) =$} &\nodata&\nodata&\nodata&\nodata & Table~\ref{likelihoods.tbl} \\
\hline

{Prior $\times$ Likelihoods}                    & {\bf 0.025} & 0.014  & 0.002  & 0.0 & Equation~\ref{BT.eqn} \\
\hline

{Normalization ($0.025 + 0.014 + 0.002 + 0.0$)} &\multicolumn{4}{c}{---0.041---} \\
\hline

{Class Posteriors}                              & {\bf 0.61} & 0.34  & 0.05  & 0.0  & $\sum_{H=1}^4 = 1$ 
\enddata
\end{deluxetable}

\subsection{Decision Rule}

Probability theory does not specify how posterior class probabilities should be used for astrophysical analyses; investigators must make that judgment themselves just as they decide whether $2\sigma$, $3\sigma$, or $4\sigma$ is the appropriate indicator of existence in Gaussian detection problems.  
We choose to adopt a class decision rule that assigns a specific class if the largest posterior probability is more than twice the next-largest posterior probability.
When no classification posterior probability stands above the others using this criterion, a source is labeled ``unclassified''.
Table~\ref{decision_rule.tbl} shows the number of sources found to have each outcome of this decision rule.
The classification details for each source are given in Table~\ref{decision_outcome.tbl}, where columns report the spatial prior probabilities, the naive Bayes posterior probabilities, and the class assignment (with ``0'' representing ``unclassified'').

\begin{deluxetable}{lcr}
\tablecaption{Decision rule outcomes for CCCP source classification \label{decision_rule.tbl}
} 
\tablewidth{0pt}
\tabletypesize{\tiny}
\tablehead{ \colhead{Class} &  \colhead{Criterion} & \colhead{Number} }
\startdata
{\rm class}=H1 (foreground star) & ${\rm Prob}(H1 \,|\, F)\; > \;2 \times \max \{{\rm Prob}(H2\,|\,F),\;{\rm Prob}(H3\,|\,F),\;{\rm Prob}(H4\,|\,F)\}$ &   716 \\ 
{\rm class}=H2 (Carina star)     & ${\rm Prob}(H2 \,|\, F)\; > \;2 \times \max \{{\rm Prob}(H1\,|\,F),\;{\rm Prob}(H3\,|\,F),\;{\rm Prob}(H4\,|\,F)\}$ & 10,714 \\ 
{\rm class}=H3 (background star) & ${\rm Prob}(H3 \,|\, F)\; > \;2 \times \max \{{\rm Prob}(H1\,|\,F),\;{\rm Prob}(H2\,|\,F),\;{\rm Prob}(H4\,|\,F)\}$ &    16 \\   
{\rm class}=H4 (extragalactic)   & ${\rm Prob}(H4 \,|\, F)\; > \;2 \times \max \{{\rm Prob}(H1\,|\,F),\;{\rm Prob}(H2\,|\,F),\;{\rm Prob}(H3\,|\,F)\}$ &   877 \\
Unclassified &  Otherwise                                                                         &  2045
\enddata
\end{deluxetable}

\begin{deluxetable}{c*{4}l*{4}lccccc}
\tablecaption{Classification quantities for CCCP sources \label{decision_outcome.tbl}
\todo{Note to referee and journal staff: the electronic table (tab4.txt) has {\bf two} sets of NH and Ltc columns that report XPHOT quantities separately for H2 low-mass sources and for other sources.  As explained in the text, readers should use values for non-H2 sources only if they have reason to believe our classification was not correct. 
}
} 
\tablewidth{0pt}
\tabletypesize{\tiny}
\tablehead{
\colhead{CXOGNC J} & \multicolumn{8}{c}{Class Probabilities}                           & \colhead{Assignment} & \colhead{$E_{\rm median}$}& \colhead{$\log \: F_{t,photon}$} &\colhead{$\log N_H$} & \colhead{$\log L_{t,c}$} \\
                 & \multicolumn{4}{c}{Spatial Prior} & \multicolumn{4}{c}{Posterior} &                 & \\
                 & \multicolumn{4}{c}{\hrulefill}    & \multicolumn{4}{c}{\hrulefill} \\  
                 & \colhead{H1} & \colhead{H2} & \colhead{H3} & \colhead{H4} & 
                    \colhead{H1} & \colhead{H2} & \colhead{H3} & \colhead{H4} \\
                 & \multicolumn{8}{c}{}                                              &                    & \colhead{(keV)} &
                 \colhead{(photon cm$^{-2}$ s$^{-1}$)}      & \colhead{(cm$^{-2}$)} & \colhead{(erg s$^{-1}$)}\\
\numberthecolumn & \numberthecolumn & \numberthecolumn & \numberthecolumn & 
\numberthecolumn & \numberthecolumn & \numberthecolumn & \numberthecolumn & \numberthecolumn &
\numberthecolumn & \numberthecolumn & \numberthecolumn & \numberthecolumn & \numberthecolumn 
\setcounter{column_number}{1}                 
}
\startdata 
  104333.37-593104.0   & 0.02 & 0.94 & 0.01 & 0.03 &    0.49 & 0.51  & 0    & 0    &  0  & 1.0 & -6.0 & \nodata & \nodata \\
  103909.94-594714.5   & 0.22 & 0.29 & 0.10 & 0.39 &    0.68 & 0.32  & 0    & 0    &  1  & 1.0 & -3.7 & \nodata & \nodata \\
  104035.22-592953.5   & 0.28 & 0.10 & 0.13 & 0.49 &    0    & 0.995 & 0    & 0    &  2  & 2.6 & -5.7 &  22.3   & 31.1  \\
  104033.22-592742.5   & 0.27 & 0.12 & 0.12 & 0.48 &    0.03 & 0.26  & 0.71 & 0    &  3  & 1.3 & -5.8 & \nodata & \nodata \\
  104323.51-592407.3   & 0.06 & 0.80 & 0.03 & 0.11 &    0    & 0.27  & 0    & 0.73 &  4  & 3.5 & -6.3 & \nodata & \nodata 
\enddata
\end{deluxetable}

%
%
%
%
%
%
%
%
%

\begin{figure}[htb]
\centering
\includegraphics[width=1.0\textwidth]{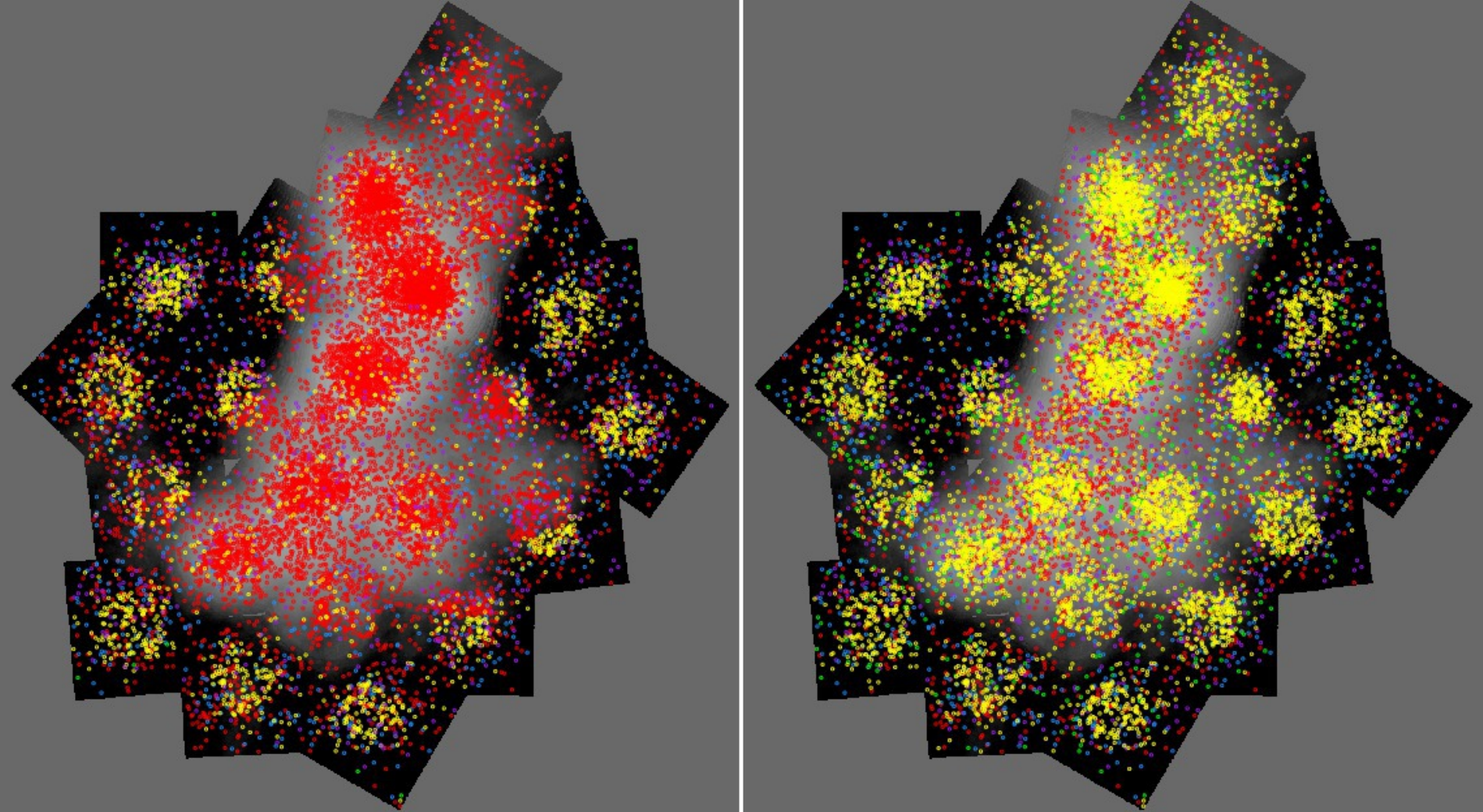}
\caption{An interpretation of posterior classification probabilities for Carina point sources derived with (left panel) and without spatial priors (right panel).
Sources are considered to be reliably classed ((H1=brown, H2=red, H3=green, H4=blue) when the largest posterior probability is more than twice the next-largest posterior probability, and are considered to be unclassified (yellow) otherwise.
The gray-scale background in both panels is the map of spatial prior probability for the H2 class (Equation~\ref{prior2.eqn}) shown in Figure~\ref{H2_prior.fig}, displayed here as white where the observed source density, and thus the H2 spatial prior, is high.
Note that the electronic version of this figure can reveal, when zoomed, detail not visible in most printed versions.
\label{classes_on_prior.fig}}
\end{figure}

The left panel of Figure~\ref{classes_on_prior.fig} shows a map of Carina sources color-coded by the choice made by our decision rule.
Within the portions of the field exhibiting high source density (gray areas in the background image) the spatial prior provides a strong preference for the H2 class (Carina members), and indeed the majority of sources are classified as such (red symbols).
The ${\sim}$2000 sources that remain unclassified (yellow symbols) presumably contain a mixture of H1, H2, H3, and H4 sources.

The right panel illustrates the important role spatial clustering plays in our classifier. 
Here we repeated the classification with the spatial priors omitted.
Ignoring clustering as a classification clue results in large numbers of unclassified sources (yellow symbols) in the known clusters, an obviously unsatisfactory result.

\subsection{X-ray Flux and Luminosity Estimates}

For the convenience of subsequent studies, Table~\ref{decision_outcome.tbl} also includes two quantities that characterize the apparent X-ray spectrum in the ``total'' energy band (0.5--8~keV).
Column (11) reproduces the median X-ray energy of the detected events, reported as column MedianEnergy\_t in Table~\tbr{1} in \citet{Broos11}.
Column (12) reports an estimate of the lowest-level calibrated photometric quantity that can be used to compare sources: apparent X-ray photon flux (photon~cm$^{-2}$~s$^{-1}$) defined \citep[][\S7.4]{Broos10}---using three quantities from Table~\tbr{1} in \citet{Broos11}---as 
\begin{quote} $F_{t,photon} \doteq$ NetCounts\_t / MeanEffectiveArea\_t / ExposureTimeNominal. \end{quote}
\citet{Broos11} use this quantity to study the spatial variation in detection completeness within the survey.

Estimation of intrinsic (absorption-corrected) flux requires two astrophysical inferences or assumptions: the absorbing column and the spectral shape of the emitting plasma.
\citet{Getman10} describe a set of plasma assumptions that are appropriate for pre-main sequence stars, a procedure for selecting the best plasma assumption based on apparent X-ray flux in the hard band (2--8~keV), a technique for inferring the absorbing column with uncertainties from the median X-ray energy statistic once a plasma has been assumed, and a set of calibrations for estimating intrinsic flux with uncertainties.
Columns (13) and (14) in Table~\ref{decision_outcome.tbl} report the resulting absorption and intrinsic X-ray luminosity values (assuming a distance of 2.3 kpc) for sources classified as pre-main sequence i.e., the H2 sample less the massive stars studied by \citet{Gagne11}.
Estimates are not available for very weak sources (those with less than 2 net counts in the hard band or less than 3 net counts in the total band).

We have also applied the \citet{Getman10} procedure to sources reported to be ``unclassified'' or likely contaminants in column (10), in anticipation that future studies may conclude that specific members of this group are instead pre-main sequence stars.  
To emphasize that this procedure is inappropriate for the {\em current} classification, we report the resulting absorption and intrinsic X-ray luminosity estimates in additional columns (not shown in the table stub) rather than in columns (13) and (14).

\section{VALIDATION OF THE CLASSIFIER \label{validation.sec}}

A supervised multivariate classification procedure is often validated by withholding a random portion of the training sample from the construction of the classifier, and later applying the decision rules to the training sample.  This procedure is called cross-validation and has a strong mathematical foundation \citep{Hastie01, Duda02}.  However, we feel this is not appropriate here as the training samples mostly come from astronomical simulations rather than true datasets.  

Recall from Table~\ref{decision_rule.tbl} that the total number of sources with a low probability of Carina membership (3654 = 716+16+877+2045) is significantly smaller than the total number of contaminants predicted from simulations (${\sim}$5,200) \citep{Getman11}.
In isolation, this fact suggests that at least ${\sim}$1,500 of the ${\sim}$10,700 sources found to be likely Carina members (H2 class) are misclassified contaminants.
However, the contaminant simulation tallies suffer from significant uncertainties, and three validation studies indicate that the H2 class does not suffer from significant residual contamination.

\begin{figure}[htb] 
\centering
\includegraphics[width=0.3\textwidth,angle=-90]{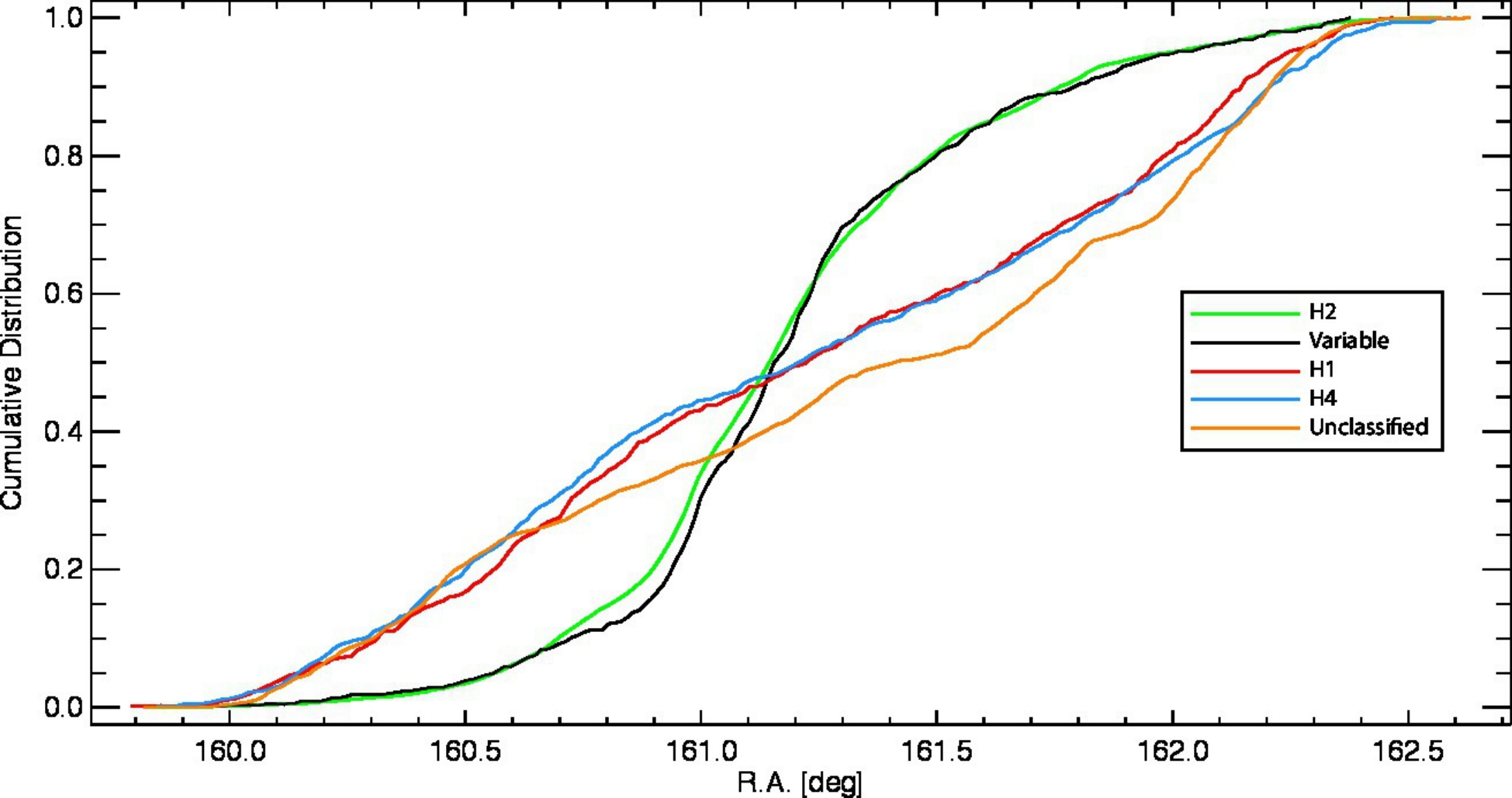}\\
\includegraphics[width=0.3\textwidth,angle=-90]{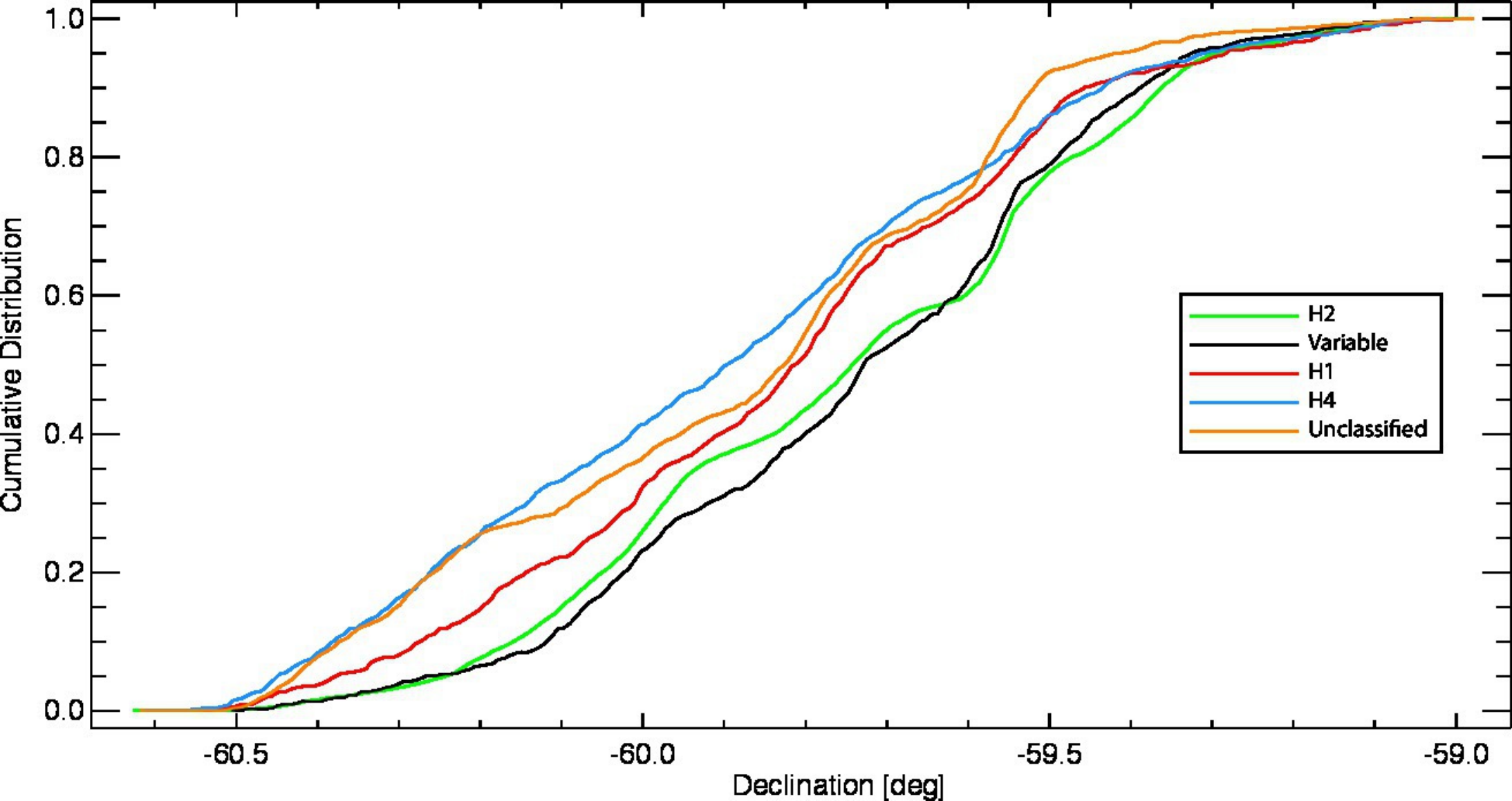}\\
\includegraphics[width=0.3\textwidth,angle=-90]{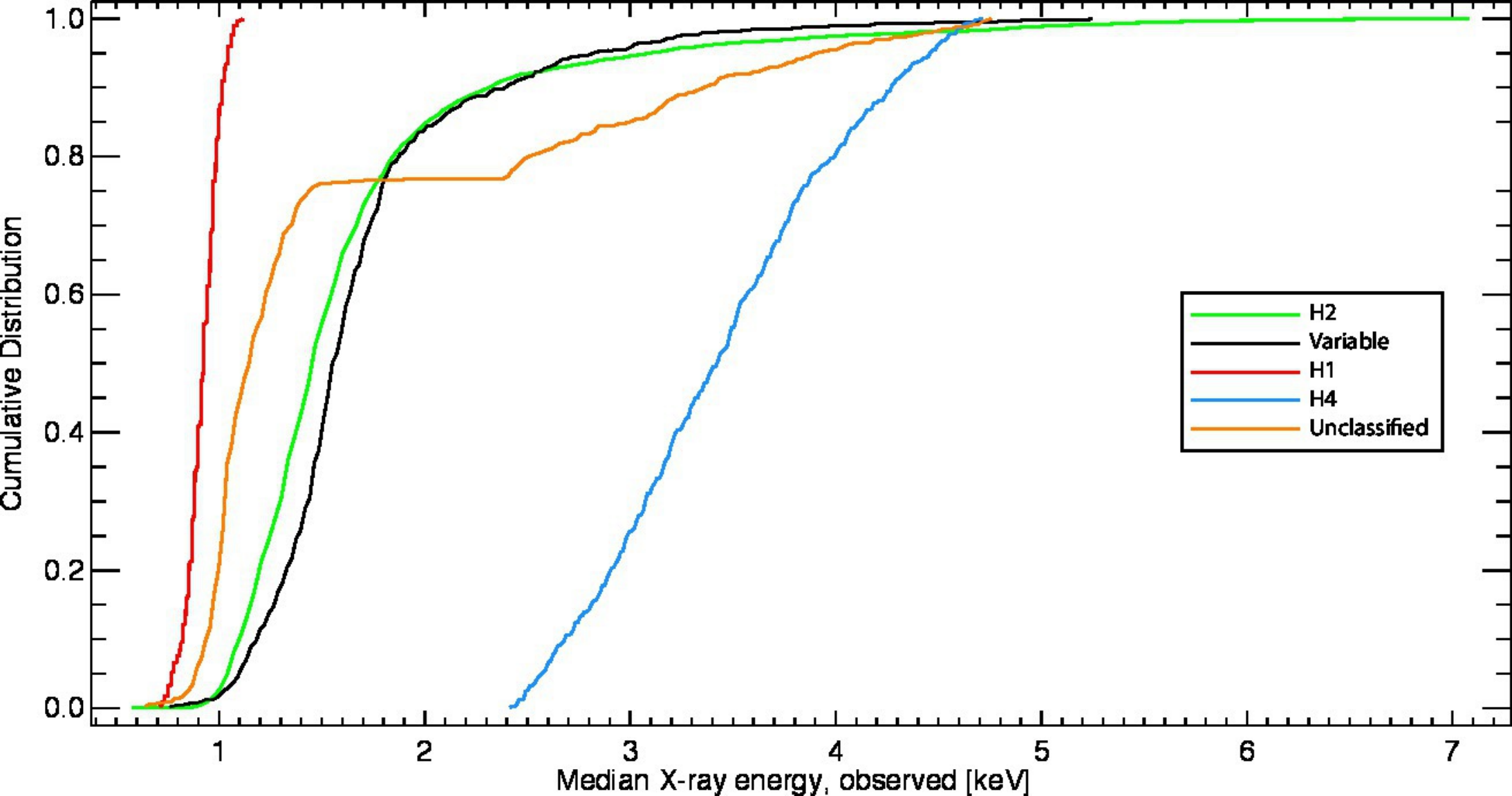}
\caption{Cumulative distributions for source position (upper and middle panels) and median X-ray energy (lower panel) for a sample of very likely Carina members (black) identified by X-ray variability, sources classified as Carina members (green), sources classified as foreground stars (red), sources classified as extragalactic (blue), and unclassified sources (orange).
\label{HiVar.fig}}
\end{figure}

Our first validation test is based on the reasonable assumption that any source exhibiting rapid X-ray variability during the CCCP observation must be a Carina member.  High-amplitude X-ray flaring is a ubiquitous characteristic of pre-main sequence stars \citep[e.g.,][]{Wolk05}, although the short CCCP exposures and low count rates preclude their detection in most members.  Figure~\ref{HiVar.fig} compares the spatial and median X-ray energy distributions of the 660 sources (black) found to be definitely variable (defined in Table~\ref{variability.tbl}) to the full set of probable Carina members (green).  These distributions follow each other accurately, and are significantly different from the distributions of sources classified as contaminants (red, blue).  This suggests that the H2 classifications have few errors, perhaps less than 5\%.  

\begin{figure}[htb]
\centering
\includegraphics[width=0.5\textwidth]{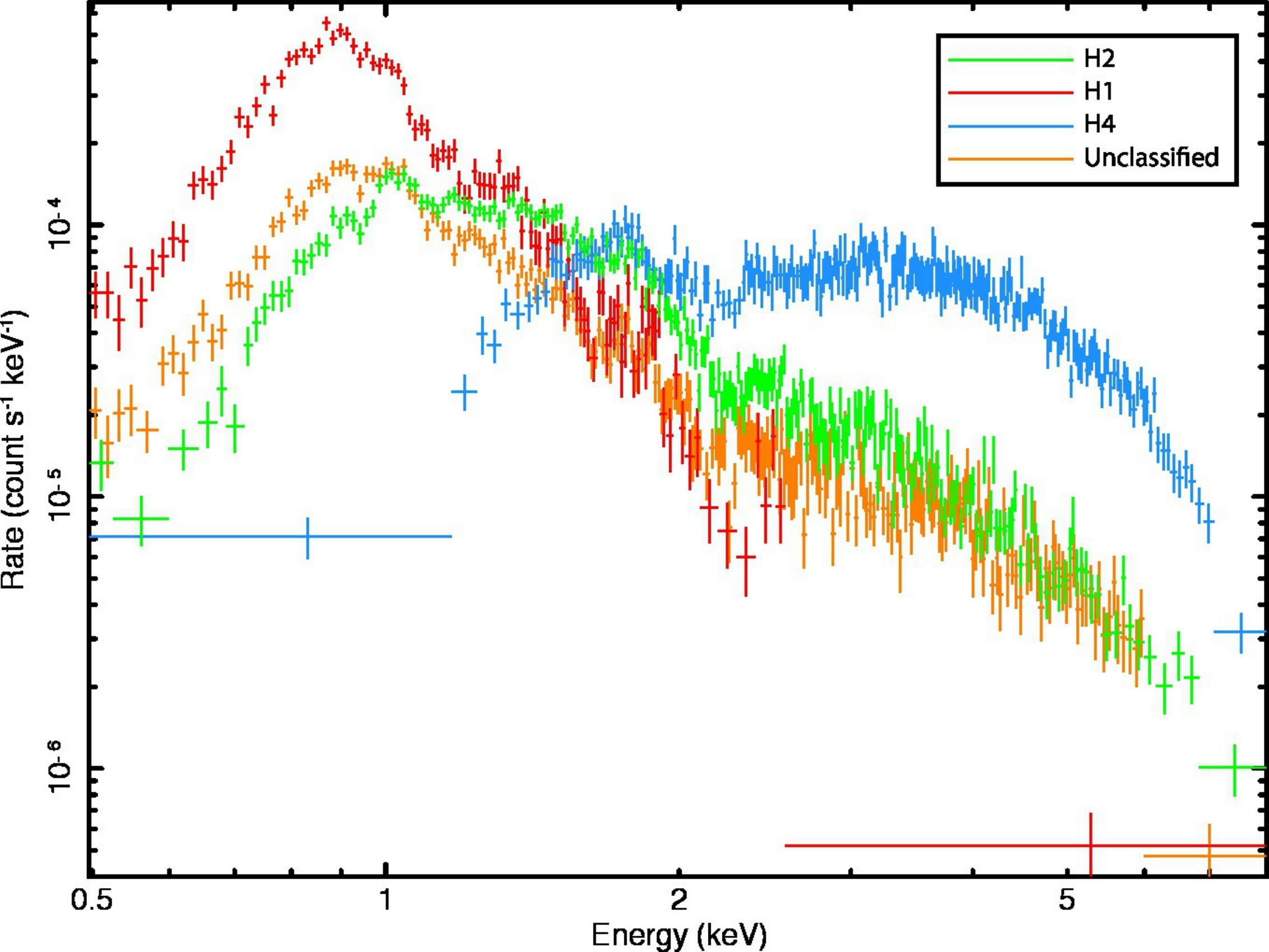}
\caption{Composite spectra for sources classified as Carina members (green), sources classified as foreground stars (red), sources classified as extragalactic (blue), and unclassified sources (orange).
The Carina member spectrum is typical for pre-main sequence stars ($N_H \simeq$ 0.2--0.5 $\times10^{22}$ cm$^{-2}$, corresponding to $A_V \simeq$ 1-3 mag; kT1 $\simeq$ 0.8 keV; kT2 $\simeq$ 3 keV).
The foreground spectrum can be fit by a much cooler thermal model ($N_H \simeq 0.2 \times 10^{22}$ cm$^{-2}$, corresponding to $A_V \simeq$ 1 mag, kT1 $\simeq$ 0.2 keV, and kT2 $\simeq$ 0.6 keV).
The extragalactic spectrum can be fit by a power-law model ($N_H \simeq 1.9 \times 10^{22}$ cm$^{-2}$, $\Gamma \simeq 1.2$), as expected for AGN.
As one might expect, the spectrum of ambiguous sources looks like a hybrid of the other three spectra. 
\label{class_spectra.fig}}
\end{figure}

Second, Figure~\ref{class_spectra.fig} shows composite (``stacked'') spectra for unclassified sources and for sources classified as Carina members, foreground stars, and extragalactic.
No composite spectrum for background stars is shown because too few sources were identified as such.
The probable Carina members show spectra similar to those seen in other pre-main sequence populations, well-modeled with plasma at temperatures around 1--3 keV \citep[e.g.,][]{Gudel09} and soft X-ray absorption equivalent to $A_V \sim$ 1--3 mag \citep{Preibisch11}.  
The sources identified as likely foreground stars have much softer, 0.2-0.6 keV \citep{Guedel97}, and less absorbed spectra.
The likely extragalactic sources have harder and more absorbed spectra as expected for AGN \citep[e.g.][]{Brandt01}.  The unclassified sources appear to be a hybrid of all classes.  These results are all qualitatively consistent with expectations for a correct classification, but it is difficult to infer quantitative error rates. We roughly estimate from these composite X-ray spectra that errors in Carina membership classification are less than 10\%. 

Third, the HAWK-I survey covers the principal clusters of Carina as well as some of the extended star forming region around the South Pillars; 6,230 CCCP sources have high-quality HAWK-I $JHK$ photometry.  \citet[][Figure 6]{Preibisch11} compare the HAWK-I near-infrared CMD for sources classified as Carina members here (class H2) with the CMD for sources classified as extragalactic contaminants and foreground field stars, as well as the unclassified sources.  The H2 source distribution appears to be a mixture of two overlapping distributions.  The great majority of CCCP/HAWK-I sources lie in a region around $13 \leq J \leq 18$ mag with reddening around $0.4 \leq \Delta(J-H) \leq 2$ mag from the expected pre-main sequence track.  
However, a small fraction (${\sim}$2\%) of the sources classified as likely Carina members lie along a more lightly reddened locus that is populated by foreground stars between $10 \leq J \leq 20$ mag \citep{Preibisch11}.  
An additional ${\sim}$4\% of sources classified as Carina members have $20 < J < 23$ mag and a wide range of reddening consistent with the locus of extragalactic sources.  

Based on these validation studies, we conclude that our sample of Carina members is probably accurate at the ${\sim}$90\% level, which we feel is quite successful given the many uncertainties involved in this analysis.


%

\section{Summary}

Simulations of Galactic field stars and extragalactic sources can provide useful tallies of the number of contaminants expected in X-ray surveys such as the CCCP.
However, the probability that an individual source is a contaminant clearly depends on its location with respect to the observed stellar clusters, its X-ray properties (e.g., variability and median X-ray energy), and its infrared properties (e.g., observed flux and infrared SED).
These observational quantities can be used to assess membership in a star forming region in various ways.
Here, we use a Naive Bayes Classifier to assign classification probabilities and adopt a decision rule to infer membership in the Carina complex for 75\% of the CCCP sources.
This sample, which is used in several CCCP studies, appears to contain only modest residual contamination.

\appendix
\section{AN ALTERNATIVE REPRESENTATION OF SPATIAL INFORMATION \label{alternative.sec}}

Inherent in the formulation of the problem into the particular data likelihoods and class priors described in \S\ref{naiveBayes.sec} is the point of view that the {\em position} of a source is not an ``observed source property'' (one of the $D_i$ quantities), and thus has no corresponding likelihood term.
This formulation asserts that we have spatially varying expectations for class fractions (represented by Bayesian priors that are functions of position) based on the clustering properties of the entire catalog, before examining data on individual sources.
This formulation is convenient because all spatial clustering information is represented in one term, the ``prior'' of Equation~\ref{posterior4.eqn}.

However, there is clearly an alternate formulation that represents clustering information in a different but equivalent way.
Under this point of view, we assert a {\em single} set of class prior probabilities averaged over the entire field,  
  \begin{eqnarray}               
    {\rm prior}_{H1} =& N1/N_{\rm CCCP} \\
    {\rm prior}_{H2} =& 1 - \frac{N1+N3+N4}{N_{\rm CCCP}} \\
    {\rm prior}_{H3} =& N3/N_{\rm CCCP} \\
    {\rm prior}_{H4} =& N4/N_{\rm CCCP} 
  \end{eqnarray}  
N1, N3, and N4 are the number of H1, H3, and H4 contaminants expected (from simulations, in our case); $N_{\rm CCCP}$ is the number of detected sources.
The form of ${\rm prior}_{H2}$ results from the requirement that any set of prior class probabilities must sum to unity.

\revise{Under this point of view, source position $\mathbf{r}$ is an observed property.
Thus, as for the other observed properties, we define four likelihoods that are probability distributions for source position conditioned on each of the hypotheses (classes):}
  \begin{eqnarray}
    \label{p_likelihood1.eqn} p(\mathbf{r} \,|\, {\rm class}=H1) =& 1 / A_{\rm CCCP} \\
    \label{p_likelihood2.eqn} p(\mathbf{r} \,|\, {\rm class}=H2) =& \frac{\rho_{\rm obs}(\mathbf{r}) -  \frac{N1+N3+N4}{A_{\rm CCCP}}}{K} \\
    \label{p_likelihood3.eqn} p(\mathbf{r} \,|\, {\rm class}=H3) =& 1 / A_{\rm CCCP} \\
    \label{p_likelihood4.eqn} p(\mathbf{r} \,|\, {\rm class}=H4) =& 1 / A_{\rm CCCP}. 
  \end{eqnarray}                                                                                      
\revise{Expressions~\ref{p_likelihood1.eqn}, \ref{p_likelihood3.eqn}, and \ref{p_likelihood4.eqn} are independent of position $\mathbf{r}$ because we have assumed that the contaminant classes (H1, H3, H4) are spatially uniform; the expressions are normalized by the area of the CCCP field, $A_{\rm CCCP}$, so that the integral of each expression across the domain of the position variable $\mathbf{r}$ (the CCCP field of view) is unity.
The numerator of Equation~\ref{p_likelihood2.eqn} is the expected position-dependent density of H2 sources (Carina members), estimated as the observed position-dependent density of detected sources ($\rho_{\rm obs}(\mathbf{r})$) minus the expected position-independent density of contaminants ($\frac{N1+N3+N4}{A_{\rm CCCP}}$).
Since the integral of Equation~\ref{p_likelihood2.eqn} across the domain of the position variable $\mathbf{r}$ must be unity, the required normalizing constant is }
  \begin{eqnarray} 
  K &=& \int_{\rm field}\: [ \rho_{\rm obs}(\mathbf{r}) -  \frac{N1+N3+N4}{A_{\rm CCCP}}]\,\mathbf{dr} \nonumber  \\
    &=& \int_{\rm field}\: \rho_{\rm obs}(\mathbf{r})\,\mathbf{dr} \: - \: \frac{N1+N3+N4}{A_{\rm CCCP}} \int_{\rm field}\,\mathbf{dr} \nonumber  \\
    &=&  N_{\rm CCCP} - (N1+N3+N4). 
  \end{eqnarray} 
The posterior class probabilities arising from this alternate formulation can be shown to be the same as those from our original formulation (Equation~\ref{posterior4.eqn}).

\acknowledgements Acknowledgments:  
We appreciate the time and useful suggestions contributed by our anonymous referee.
This work is supported by Chandra X-ray Observatory grant GO8-9131X (PI:  L.\ Townsley) and by the ACIS Instrument Team contract SV4-74018 (PI:  G.\ Garmire), issued by the {\em Chandra} X-ray Center, which is operated by the Smithsonian Astrophysical Observatory for and on behalf of NASA under contract NAS8-03060.
M.S.P. is supported by an NSF Astronomy and Astrophysics Postdoctoral Fellowship under award AST-0901646.
This publication makes use of data products from the Two Micron All Sky Survey, which is a joint project of the University of Massachusetts and the Infrared Processing and Analysis Center/California Institute of Technology, funded by the National Aeronautics and Space Administration and the National Science Foundation.
The HAWK-I near-infrared observations were collected with the High Acuity Wide-field K-band Imager instrument \citep{Kissler-Patig08} on the ESO 8-meter Very Large Telescope (VLT) at Paranal Observatory, Chile, under ESO programme 60.A-9284(K).
This work is based in part on observations made with the {\em Spitzer Space Telescope}, which is operated by the Jet Propulsion Laboratory, California Institute of Technology under a contract with NASA.

{\em Facilities:} \facility{CXO (ACIS)}, \facility{Spitzer (IRAC)}, \facility{VLT:Yepun (HAWK-I)}, \facility{FLWO:2MASS ()}, \facility{CTIO:2MASS ()}


\todo{Note to referee and journal staff: to reduce cost Figure \ref{prior_and_posterior.fig} could appear in B\&W in print if it cannot share a page with another color figure.}

\end{document}